%% file: main.tex
\pdfoutput=1

\documentclass[iop]{emulateapj}

\usepackage[normalem]{ulem}
\usepackage{color}

\usepackage{textcomp}

\newcommand{\kgsdel}[1]{{}}

\usepackage[export]{adjustbox}
\usepackage{graphics}
\usepackage{hyperref}

\usepackage[figuresright]{rotating}

\hypersetup{
    pdfauthor=Stefan Laos,
    pdftitle=Work in Progress,
    colorlinks=true,
    citecolor=blue,
    linkcolor=black,
    urlcolor=cyan}

\shorttitle{Keck Protostar Paper}
\shortauthors{Laos, Greene, Najita \& Stassun}

\DeclareUnicodeCharacter{2212}{-}

\begin{document}

%% LaTeX will automatically break titles if they run longer than
%% one line. However, you may use \\ to force a line break if
%% you desire.

% \title{Characterizing the inner disk regions of Class 0 protostars with near-infrared spectroscopy}

\title{The near stellar environment of Class 0 protostars: a first look with near-infrared spectroscopy}

%% Use \author, \affil, and the \and command to format
%% author and affiliation information.
%% Note that \email has replaced the old \authoremail command
%% from AASTeX v4.0. You can use \email to mark an email address
%% anywhere in the paper, not just in the front matter.
%% As in the title, use \\ to force line breaks.

\author{Stefan Laos\altaffilmark{1}, Thomas P.\ Greene\altaffilmark{2}, Joan R. Najita\altaffilmark{3}, and Keivan G.\ Stassun\altaffilmark{1}}

\altaffiltext{1}{Department of Physics and Astronomy, Vanderbilt University, Nashville, TN 37235, USA}
\altaffiltext{2}{NASA Ames Research Center
Space Science and Astrobiology Division M.S. 245-6 Moffett Field, CA 94035, USA}
\altaffiltext{3}{National Optical Astronomy Observatory, 950 N. Cherry Avenue, Tucson, AZ 85719, USA}

%\showthe\columnwidth

% Abstract
\input{Abstract.tex}

% Introduction
\input{Introduction.tex}

\input{Obs_data_reduction.tex}

\input{Analysis.tex}

\input{Discussion.tex}

\input{Summary_Conclusion.tex}

\input{Acknowledgements.tex}

% Bibliography
\bibliographystyle{aasjournal}
\bibliography{Keck_protostar.bib}
%\nocite{*}

\appendix
\input{SED_fitting.tex}

% \begin{figure*}[!htp]
%     \includegraphics[width=\linewidth]{Figures/CompositeH2.pdf}
%     \caption{Absolute look at fluxed \textit{K}--band spectra of Class 0 protostar sample. These spectra have not been continuum subtracted.  We note the continuum of Perseus 21 is not detected. HOPS 44 is also not included for clarity. }
%     \label{fig:compositefull}
% \end{figure*}
\clearpage

\begin{sidewaystable}
\centering
\caption{Emission Line Equivalent Widths}
\label{tab:EWtable}
\begin{tabular}{ccccccccccccccc}
\hline \hline
Source & 1-0 S(2) & 2-1 S(3) & 1-0 S(1) & 2-1 S(2) & 3-2 S(3) & 1-0 S(0) & 2-1 S(1) & Br y & CO (2-0) & CO (3-1) & CO (4-2) \\
 & EW (\AA) & EW (\AA) & EW (\AA) & EW (\AA) & EW (\AA) & EW (\AA) & EW (\AA) & EW (\AA) & EW (\AA) & EW (\AA) & EW (\AA) \\
\hline
Per-emb 26 & 57.3 ± 1.1 & 29.4 ± 1.0 & 110.0 ± 1.0 & 8.3 ± 1.0 & 5.5 ± 1.0 & 52.4 ± 1.1 & 16.6 ± 1.1 & $\leq$1.3\tablenotemark{a} & 25.5 ± 1.6 & 29.6 ± 1.6 & 27.6 ± 1.6 \\
Per-emb 25 & 2.1 ±  0.2 & 1.9 ± 0.2 & 1.7 ± 0.2 & 0.9 ± 0.2 & $\leq$0.2\tablenotemark{a} & 3.3 ± 0.2 & $\leq$0.2\tablenotemark{a} & 4.5 ± 0.3 & 25.1 ± 0.3 & 20.3 ± 0.3 & 19.2 ± 0.3 \\
Per-emb 28 & 3.9 ± 0.1 & 2.0 ± 0.1 & 6.8 ± 0.1 & 1.1 ± 0.1 & $\leq$0.1\tablenotemark{a} & 1.4 ± 0.1 & 0.8 ± 0.1 & 2.9 ± 0.1 & 12.3 ± 0.1 & 8.6 ± 0.1 & $\leq$0.1\tablenotemark{a} \\
Per-emb 8 & 51.0 ± 0.3 & 9.5 ± 0.3 & 114.1 ± 0.3 & 5.7 ± 0.3 & 2.9 ± 0.4 & 35.0 ± 0.3 & 6.3 ± 0.4 & 6.9 ± 0.4 & $\leq$0.5\tablenotemark{a} & $\leq$0.5\tablenotemark{a} & $\leq$0.5\tablenotemark{a} \\
HOPS 32 & 55.4 ± 0.3 & 14.1 ± 0.3 & 149.8 ± 0.3 & 6.2 ± 0.3 & 3.2 ± 0.3 & 25.5 ± 0.3 & 12.5 ± 0.3 & 4.5 ± 0.4 & 25.9 ± 0.4 & 17.5 ± 0.4 & 11.0 ± 0.4 \\
\hline
\end{tabular}
\footnotetext[0]{All EWs correspond to emission features. We report their absolute values in this table. These EWs do not account for emission extended from the continuum.}
\footnotetext[1]{We report a 3$\sigma$ upper limit for this undetected emission line.}
\end{sidewaystable}

\begin{table*}
\centering
\caption{Emission Line Derivations}
\label{tab:Analysistable}
\begin{tabular}{cccccccc}
\hline \hline
Source & CO (2-0) FWHM\tablenotemark{a, b} & Br y FWHM\tablenotemark{a} & 1-0 S(1) FWHM\tablenotemark{a, c} & 1-0 S(1) Offset & 1-0 S(1)/2-1 S(1) & 1-0 S(1) extent & Av$_{v}$\tablenotemark{e} \\
 & km/s & km/s & km/s & km/s &  & (") & mag \\
\hline
Per-emb 26 & 352 &  & 119 & 68 & 9.4 & ~13 & 18.7\tablenotemark{d} \\
Per-emb 25 & 201 & 307 & 129 & 58 &  &  & 9.3 \\
Per-emb 21 &  &  & 26 & -46 & 19.6 & ~10 & 36.4 \\
Per-emb 28 & 202 & 244 & 42 & 75 & 12.5 &  & 4.7 \\
Per-emb 8 &  & 219 & 139 & 68 & 11.3 & ~5 & 20.9 \\
HOPS 32 & 239 & 237 & 50 & -15 & 9.4 & ~8 & 23.9 \\
\hline
\end{tabular}
\footnotetext[1]{The intrinsic instrumental line width of 120 km s$^{−1}$ has been removed in quadrature.}
\footnotetext[2]{These FWHMs are derived via the width of our Gaussian filter in our synthetic spectra comparison (Section~\ref{subsec_COmodeling}).}
\footnotetext[3]{These FWHMs are subject to the variable spatial extents of our objects (Table~\ref{tab:journalobs}) and are broadly overestimated. (Section~\ref{subsec_ratios}).}
\footnotetext[4]{Poorly constrained due to significant variability in the IRAC photometry of Per-emb 26 (Appendix Section~\ref{subsec_bestfit}).}
\footnotetext[5]{Best-fit values from SED model analysis
(Appendix Section~\ref{sec:SED_fitting}). We note these derivations are very approximate, likely lower, estimates given the complicated physical nature of Class 0 SEDs (see Section~\ref{subsec_luminosities}, Appendix Section A).}
\end{table*}

\begin{table*}
\centering
\caption{Position Angle Comparison}
\label{tab:PAtable}
\begin{tabular}{cccc}
\hline \hline
Object & Slit PA & Outflow PA & Reference\tablenotemark{a} \\
\hline
Per-emb 26 & 4 & 162, 165 & 1, 2 \\
Per-emb 25, Oct 12/13 & 0, 0 & 104, 105 & 1, 2 \\
Per-emb 21 & 320 & 48 & 3 \\
Per-emb 28 & 150 & 112 & 3 \\
Per-emb 8, Oct 13/14 & 313, 133 & 314 & 4 \\
\hline 
\end{tabular}
\footnotetext[1]{Reference IDs: \citet{2017ApJ...846...16S} (1), \citet{2008MNRAS.387..954D} (2), \citet{2016ApJ...820L...2L} (3), \citet{2018ApJ...867...43T} (4).}
\end{table*}

\begin{table*}
\centering
\caption{Emission Incidence Statistics}
\label{tab:Emissionstats}
\begin{tabular}{ccc}
\hline \hline
Emission Feature & Class 0s\tablenotemark{a} & Class 1s\tablenotemark{b} \\
\hline
Br $\gamma$ & 4/6 $\cong$ 67\% & 34/52 $\cong$ 65\% \\
H$_{2}$ & 6/6 = 100\% & 23/52 $\cong$ 44\% \\
CO & 4/6 $\cong$ 67\% & 8/52 $\cong$ 15\% \\
\hline
\end{tabular}
\footnotetext[1]{This Class 0 sample constitutes the 5 continuum-detected Class 0s from this work (Per-emb 8, Per-emb 25, Per-emb 26, Per-emb 28, and HOPS 32) and Serpens S68N (\citealt{2018ApJ...862...85G}).}
\footnotetext[2]{This Class 1 sample corresponds to the Class 1s observed in nearby dark clouds (\citealt{2005AJ....130.1145D}).}
\end{table*}
\clearpage

\begin{figure*}[!htp]
    \includegraphics[width=\textwidth]{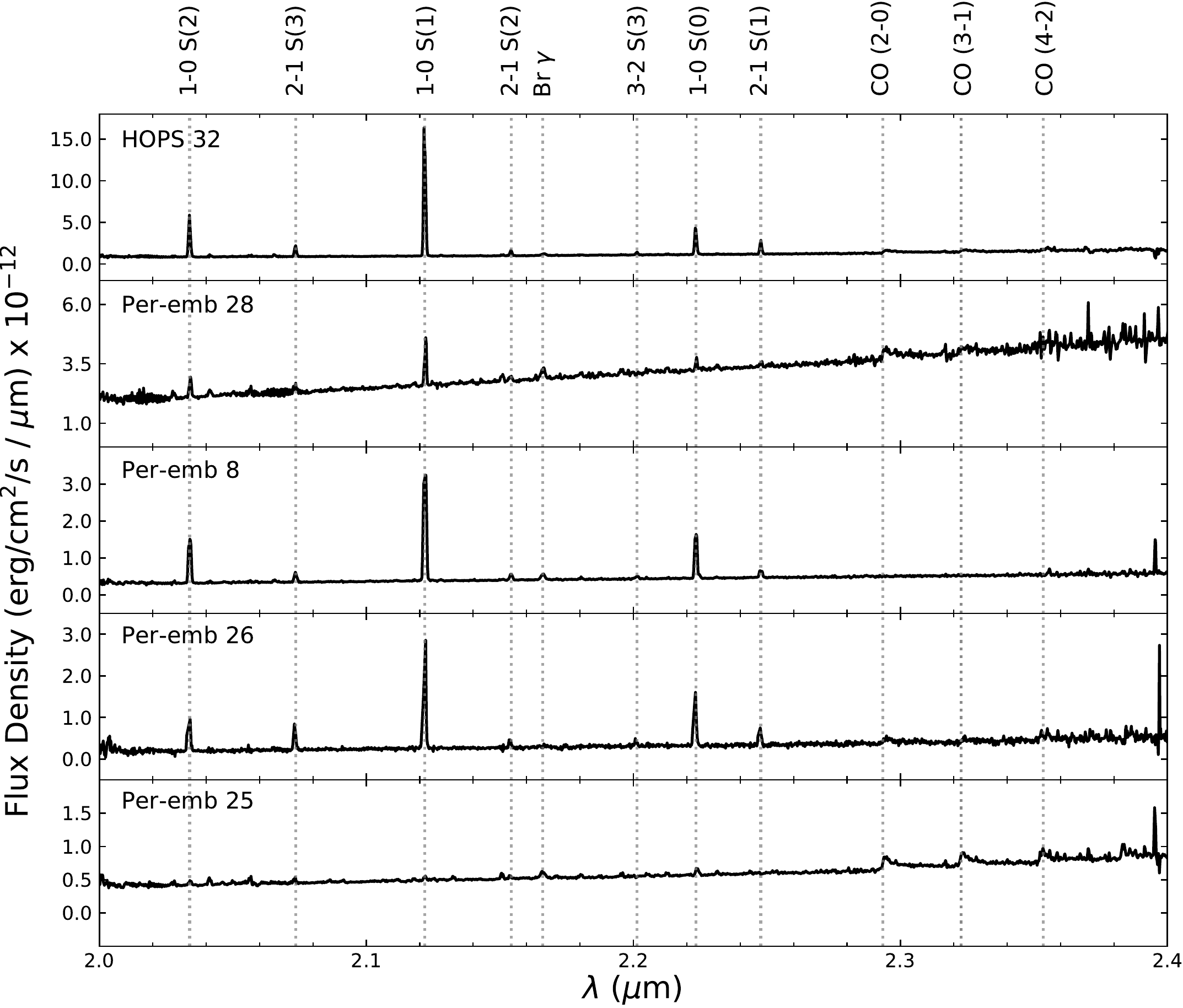}
    \caption{\textit{K}--band spectra of our Class 0 protostar sample, excluding Per-emb 21 and HOPS 44 (see Section~\ref{sec:analysis}). We identify the major atomic and molecular emission features present across the spectra (numerous H$_{2}$ emission line transitions, Br $\gamma$ emission, and CO band emission). We refer the reader to Table~\ref{tab:EWtable} for a complete census of line feature detection per individual source.}
    \label{fig:compositespec}
\end{figure*}

% \begin{figure*}[h]
%     \includegraphics[width=\linewidth]{Figures/Br_gamma_look_vel.pdf}
%     \caption{Br $\gamma$ line profiles for our Keck sample. Spectra have been continuum normalized. The light grey dotted line indicates line center while the darker lines  indicate +- 60 km/s, given our nominal velocity resolution of $\sim$120 km/s.}
%     \label{fig:Br_gamma_profiles}
% \end{figure*}

% \begin{figure*}[!htp]
%     \includegraphics[width=\linewidth]{Figures/CompositeCO.pdf}
%     \caption{Zoomed-in look at detected 2-0 and 3-1 CO bands for our Class 0 protostar sample. The spectra have been shifted into the systemic rest frame, continuum normalized, and vertically shifted for clarity. Dotted lines indicate the rest wavelengths of the CO 2-0 and 3-1 bandheads from left to right.}
%     \label{fig:compositeCO}
% \end{figure*}

\begin{figure*}[h]
    \centering
    \includegraphics[width=\textwidth]{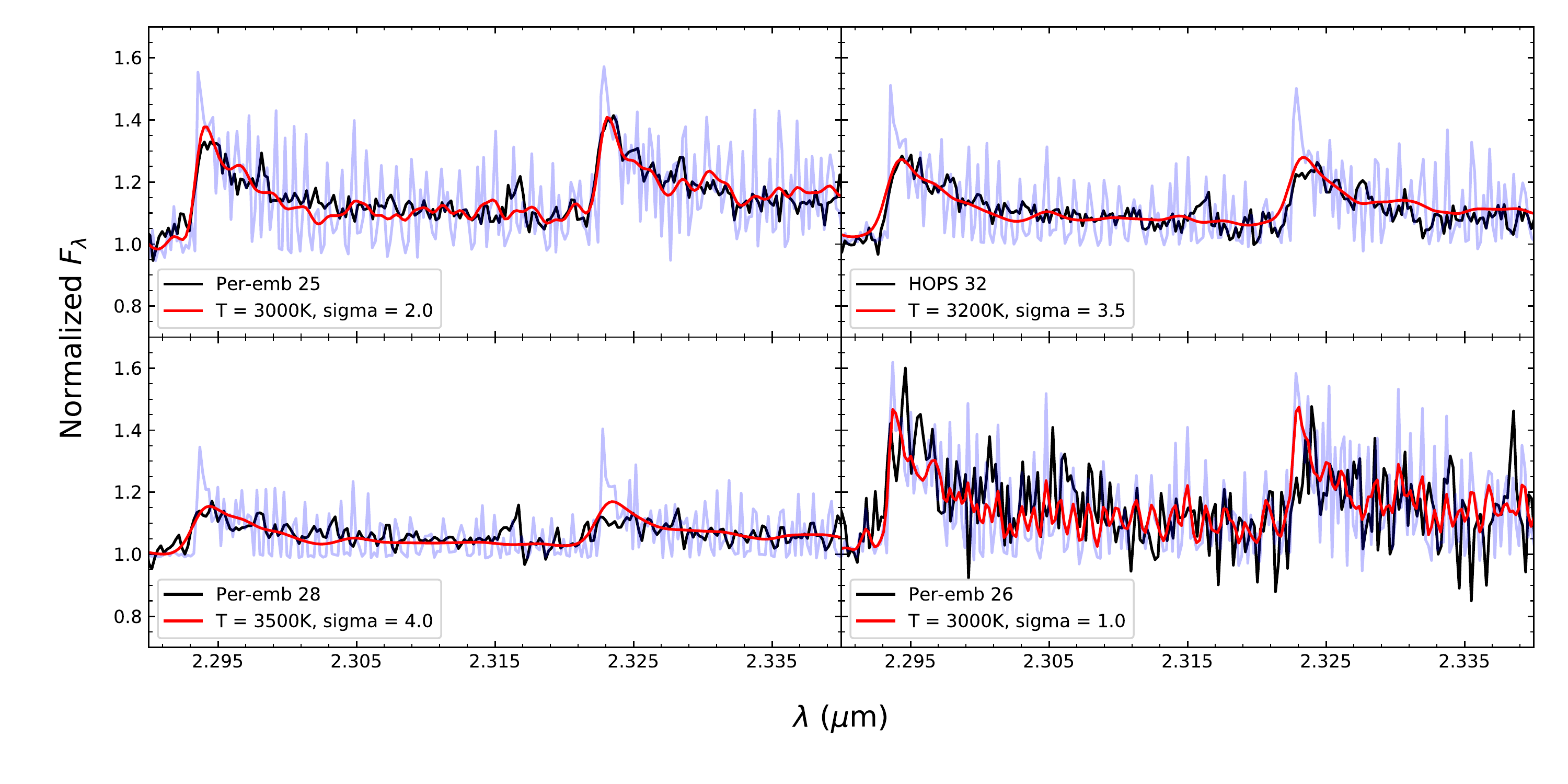}
    \caption{Our observed CO v=2–0 emission bands (black) overplotted with that of inverted PHOENIX model spectra before (light blue) and after (red) smoothing with a Gaussian filter (Section~\ref{subsec_COmodeling}). The reasonable match between our data and model spectra demonstrate that spectrally broadened CO emission is detected in these four Class 0 objects.}
    \label{fig:COmodel_samplefits}
\end{figure*}

\begin{figure*}[!htp]
    \includegraphics[width=\textwidth]{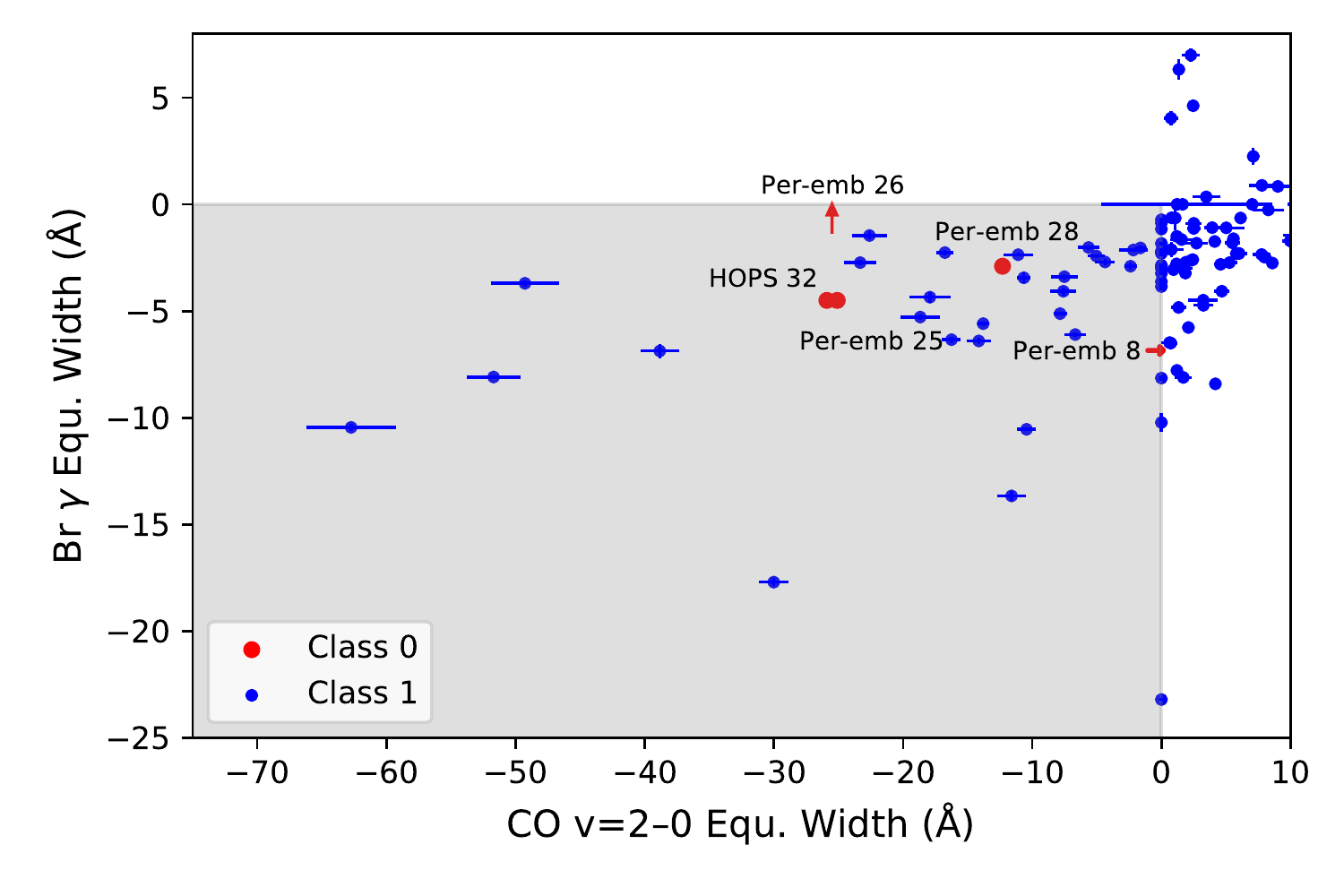}
    \caption{Br $\gamma$ equivalent widths vs. the CO v=2–0 band equivalent widths. Red points correspond to observed Class 0 sources (this work) while blue points correspond to observed Class I sources \citep{2010AJ....140.1214C}. The shaded region demarcates sources with observed emission in both features. Undetected lines are represented with arrows, corresponding to 3$\sigma$ upper limits. We find the broad distribution of these EWs appear to overlap between Class 0 and Is (Section~\ref{subsec_lineprofiles}).}
    \label{fig:Connelley_CO_Bry}
\end{figure*}

\begin{figure*}[h]
    \centering
    \includegraphics[trim=20 10 0 0, clip, width=\textwidth]{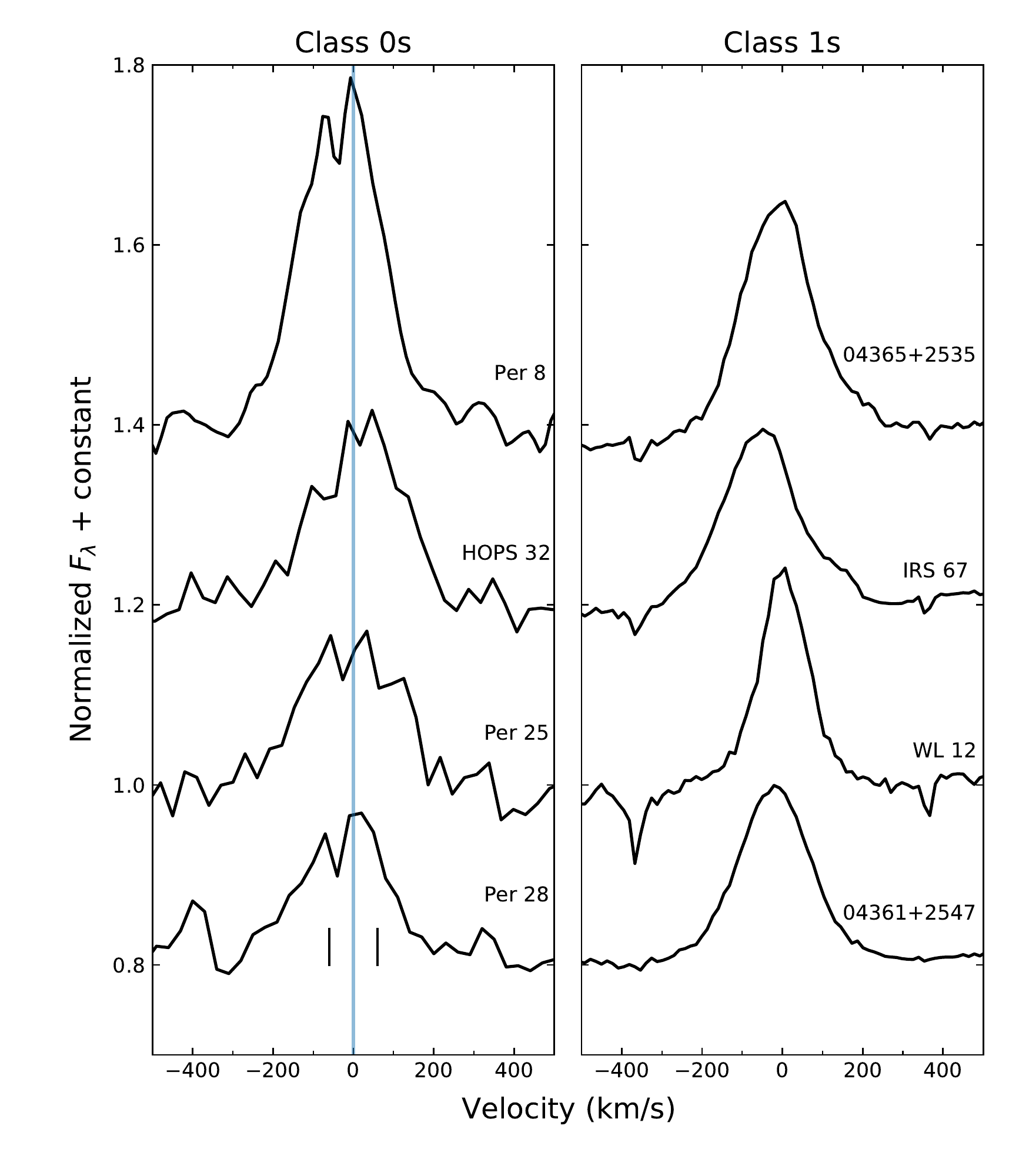}
    \caption{Comparison of observed Br $\gamma$ line profile between our Class 0s (left) and the Class Is from \citet{2005AJ....130.1145D} (right). Spectra have been normalized with a linear fit to the continuum and shifted into the systemic reference frame (Section~\ref{subsec_lineprofiles}). We overplot a bisecting line in blue to roughly quantify the relative blue and red shifted flux contributions visually. The black ticks represent the approximate velocity resolution of our data ($\sim$120 km/s). The high resolution spectra of the Class Is (R=18,000) has been downgraded to match our moderate MOSFIRE resolution (R=2,400). In Section~\ref{subsec_magnetosphere}, we argue the similarity between our Class 0 line profiles and that of Class 1s hints towards the presence of a magnetosphere at the Class 0 stage.}
    \label{fig:Br_gamma_profiles}
\end{figure*}

\begin{figure*}[!htp]
    \includegraphics[width=\linewidth]{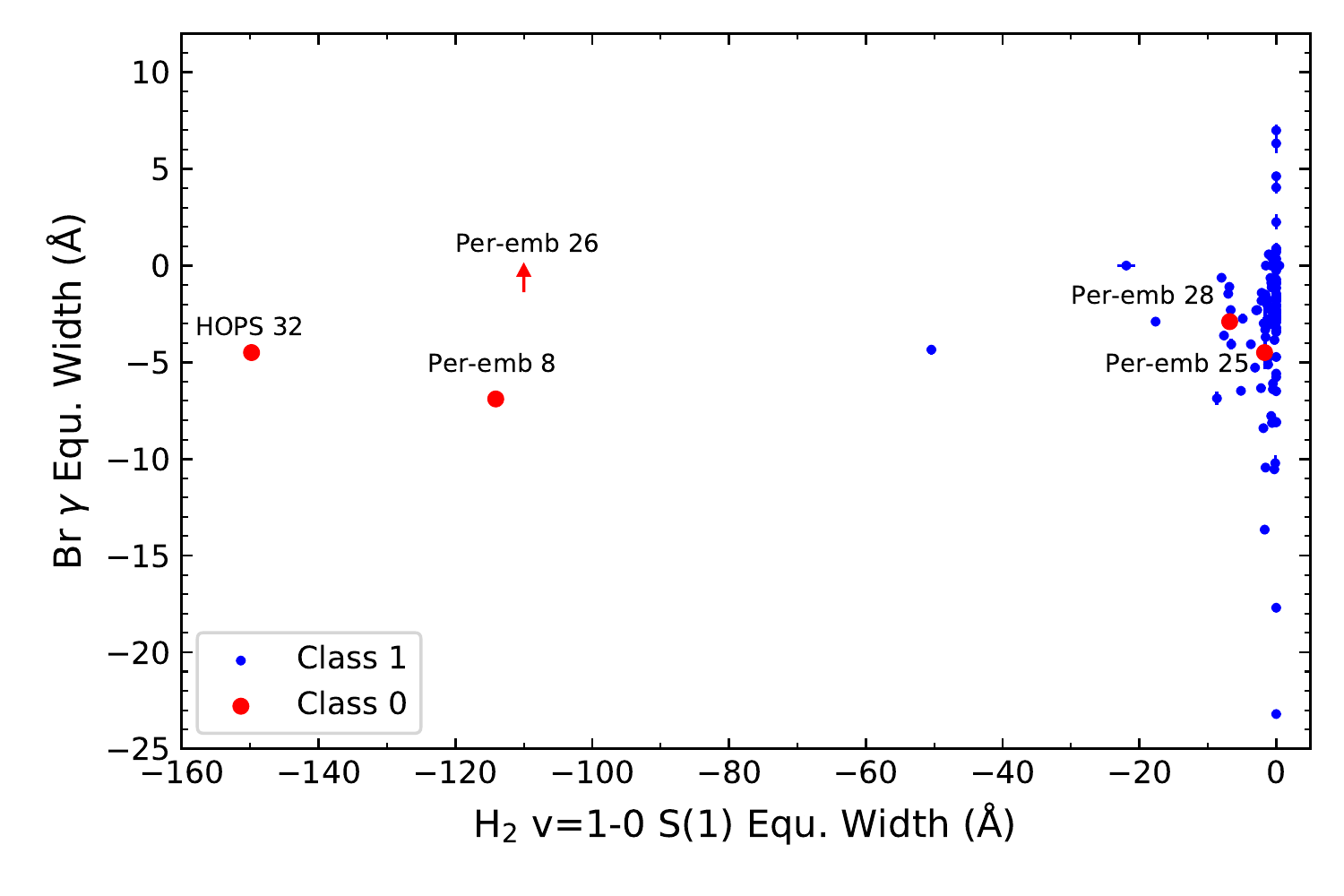}
    \caption{Br $\gamma$ equivalent widths vs. H$_{2}$ v=1-0 S(1) equivalent widths. Red points correspond to observed Class 0 sources (this work) while blue points correspond to observed Class I sources \citep{2010AJ....140.1214C}. Undetected lines are represented with arrows, corresponding to 3$\sigma$ upper limits. We find some of our Class 0s exhibit significantly higher H$_{2}$ v=1-0 S(1) EWs than this sample of Class Is (Section~\ref{subsec_ratios}).}
    \label{fig:Connelley_H2_Bry}
\end{figure*}

\begin{figure*}[h]
    \centering
    \includegraphics[width=\textwidth]{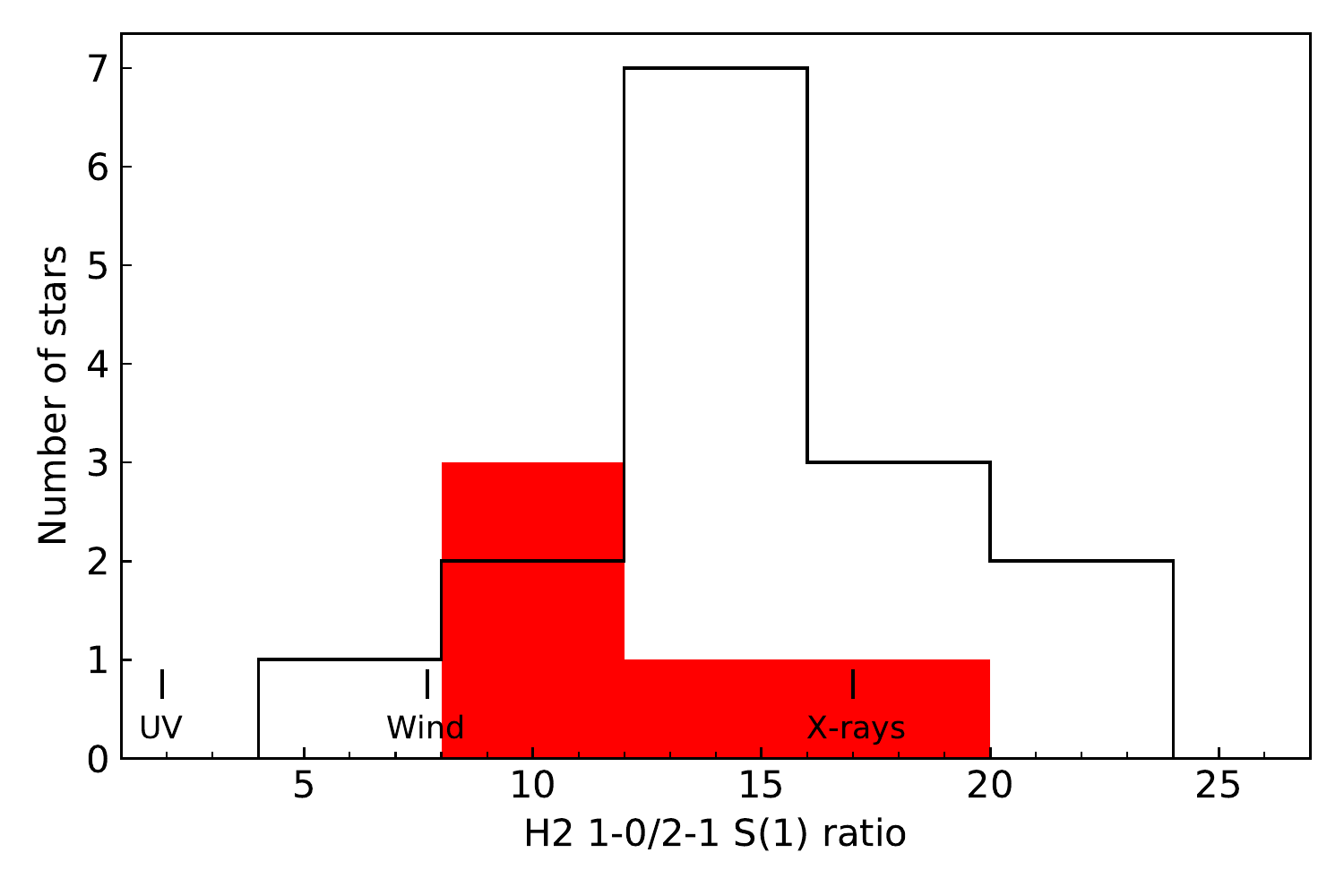}
    \caption{Histogram of H$_{2}$ $\nu$ = 1–0/2-1 S(1) line ratios between our Class 0s (red) and the Class Is observed in \citet{2010ApJ...725.1100G} (black). We find the line ratios of our Class 0s argue in favor of shocks in a wind or X-rays as the most likely  H$_{2}$ source excitation mechanism.}
    \label{fig:H2lineratiocomp}
\end{figure*}

\begin{figure*}[!htp]
    \centering
    \includegraphics[width=1.0\textwidth]{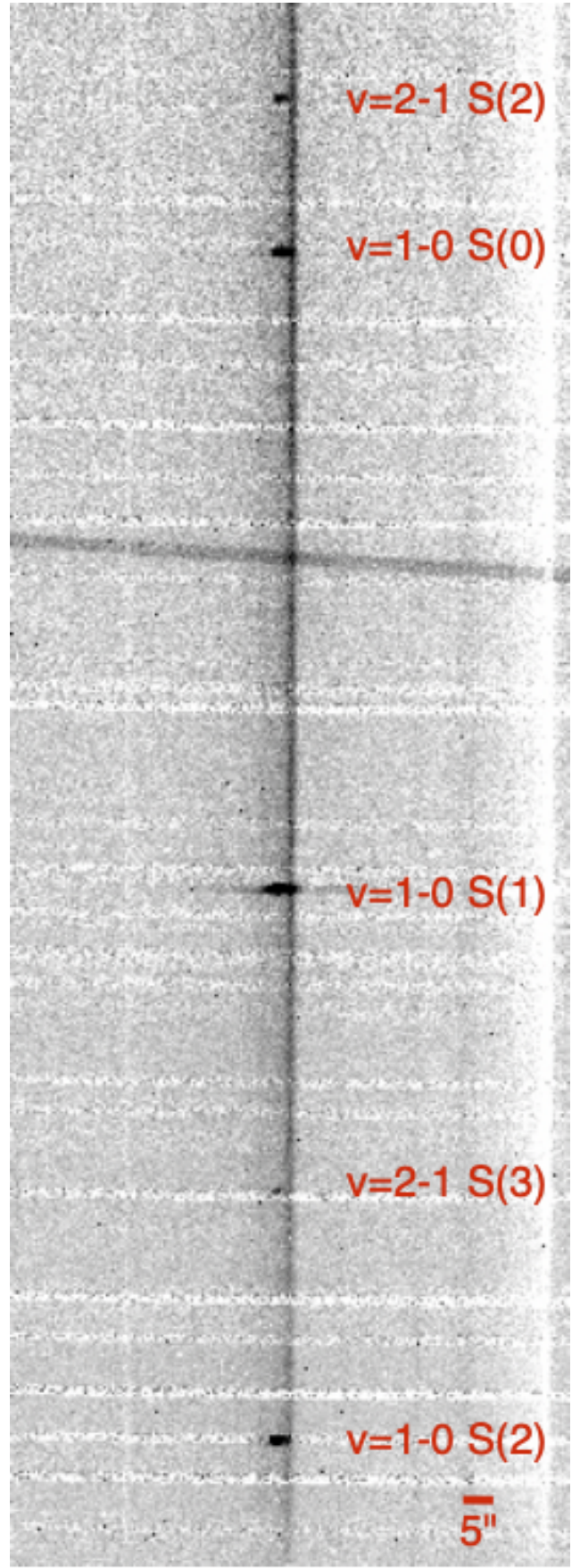}
    \caption{An example of the extended H$_{2}$ emission seen in our reduced 2d spectral images. The frame cut is centered on the observed continuum of Per-emb 8 from a single 120s exposure observed on UT date 2019 October 14. We label the individual features and discuss the slit position angle (PA) alignment with the PA of known outflows in Section~\ref{subsec_extended}.}
    \label{fig:Per8extend}
\end{figure*}

\begin{figure*}[h]
    \centering
    \includegraphics[width=\textwidth]{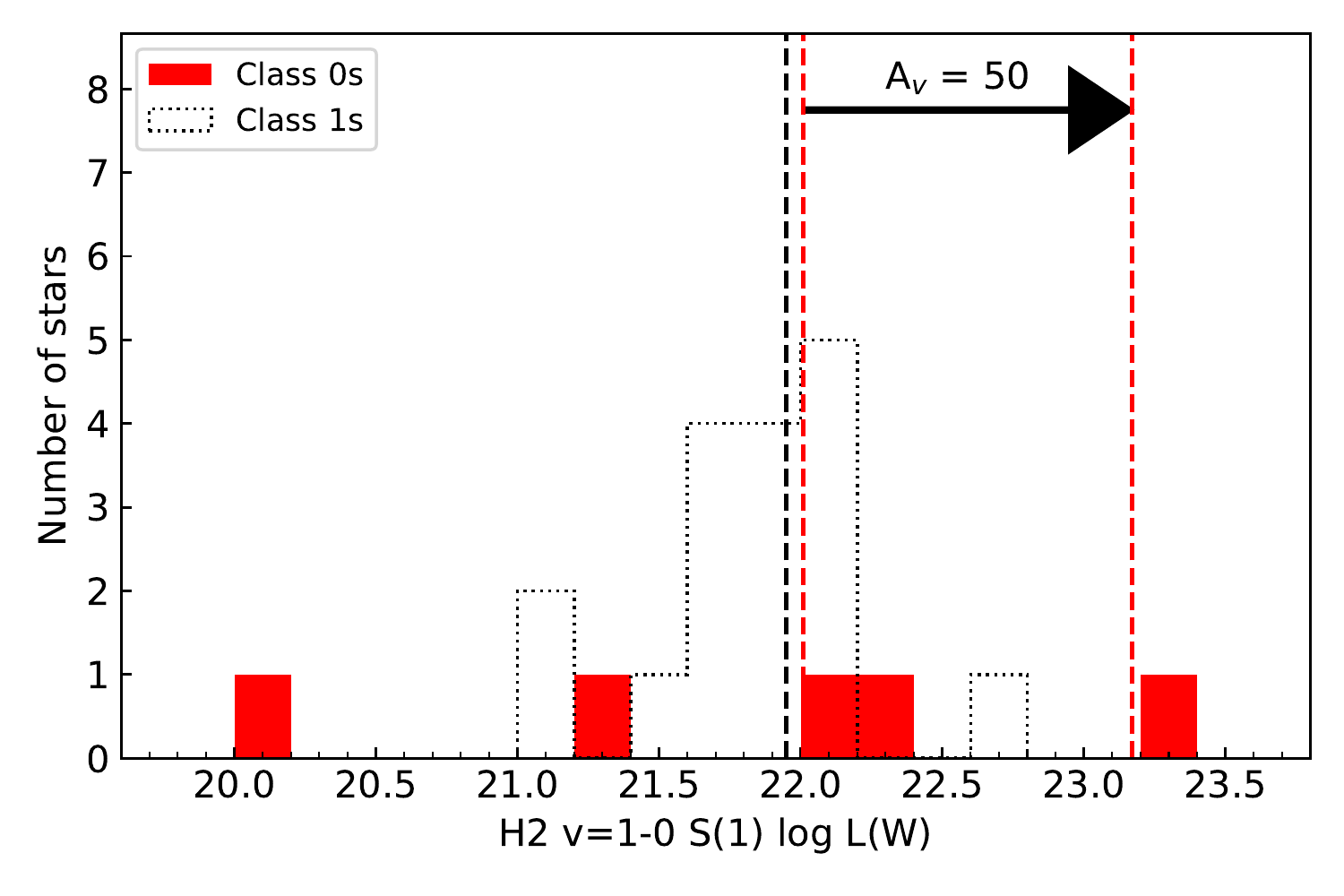}
    \caption{Histogram of H$_{2}$ $\nu$ = 1–0 S(1) line luminosities between our Class 0s (red) and the Class Is observed in \citet{2010ApJ...725.1100G} (black). We note our Class 0 line luminosities have high uncertainties (Section~\ref{subsec_luminosities}); the visual extinctions derived for de-reddening are likely underestimated given the complicated physical nature of our Class 0s (e.g. non-symmetrical envelopes, scattered light). We indicate the median H$_{2}$ $\nu$ = 1–0 S(1) line luminosity for the Class 1 sample (vertical dashed black line). In red, we also indicate this median value for our Class 0 line luminosities (vertical red dashed line) after de-reddening by the extinction estimates from our SED fitting (left) or by an effective extinction of 50 magnitudes for all of our Class 0 sources (right). More precise Class 0 extinction estimates are needed to determine the extent to which the H$_{2}$ line luminosities of Class 0 and I sources differ.}
    \label{fig:LogH2lumcomp}
\end{figure*}

\begin{figure*}[h]
    \centering
    \includegraphics[width=\textwidth]{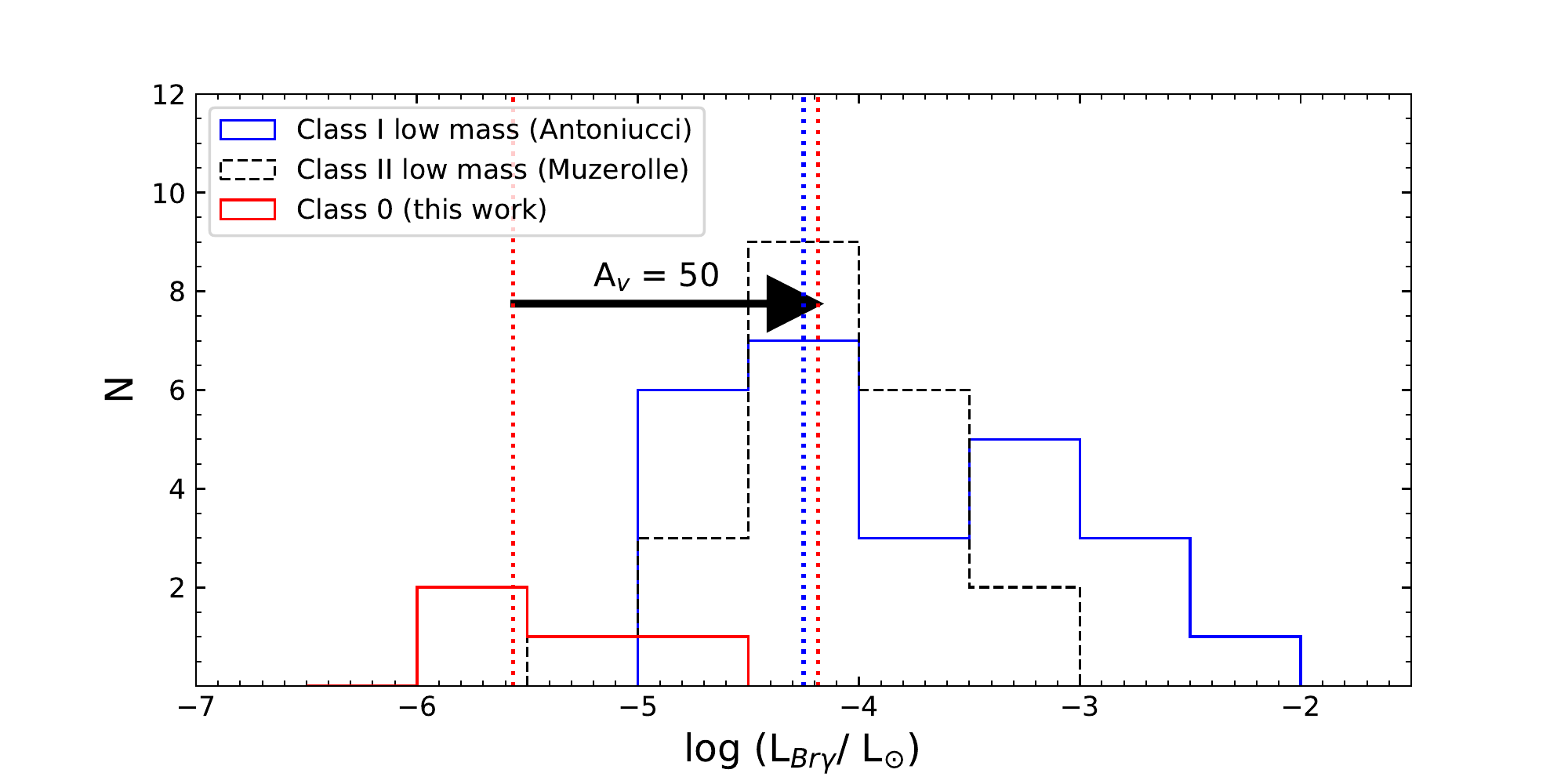}
    \caption{Histogram of Br $\gamma$ line luminosities between our Class 0s (red), low mass Class Is (\citealt{2008A&A...479..503A}) (blue), and low (\citealt{1998ApJ...492..743M}) mass Class IIs (black). As in Figure~\ref{fig:LogH2lumcomp}, we note our Class 0 line luminosities have high uncertainties (Section~\ref{subsec_luminosities}). We indicate the median Br $\gamma$ line luminosity for the Class 1 and II samples (vertical dashed blue line). In red, we also indicate this median value for our Class 0 line luminosities (vertical red dashed line) after de-reddening by the foreground extinction estimates from our SED fitting (left) or by an effective extinction of 50 magnitudes for all of our Class 0 sources (right). %Overall, it remains unclear the extent to which the line luminosities of these two populations differ until more precise Class 0 extinction estimates can be derived.
    More precise Class 0 extinction estimates are needed to determine the extent to which the Br$\gamma$ luminosities of Class 0 and I sources differ.}
    \label{fig:Brylumcomp}
\end{figure*}

% \begin{figure*}[!htp]
%     \includegraphics[width=\linewidth]{Figures/COvsBry_luminosity.pdf}
%     \caption{Estimated CO luminosity vs. Br $\gamma$ luminosity for our Keck sample and the young stellar objects from cite Carr 1989.}
%     \label{fig:CO_Brgamma}
% \end{figure*}

\begin{figure*}[htp]
    \includegraphics[trim=0 80 0 0, clip, width=\textwidth]{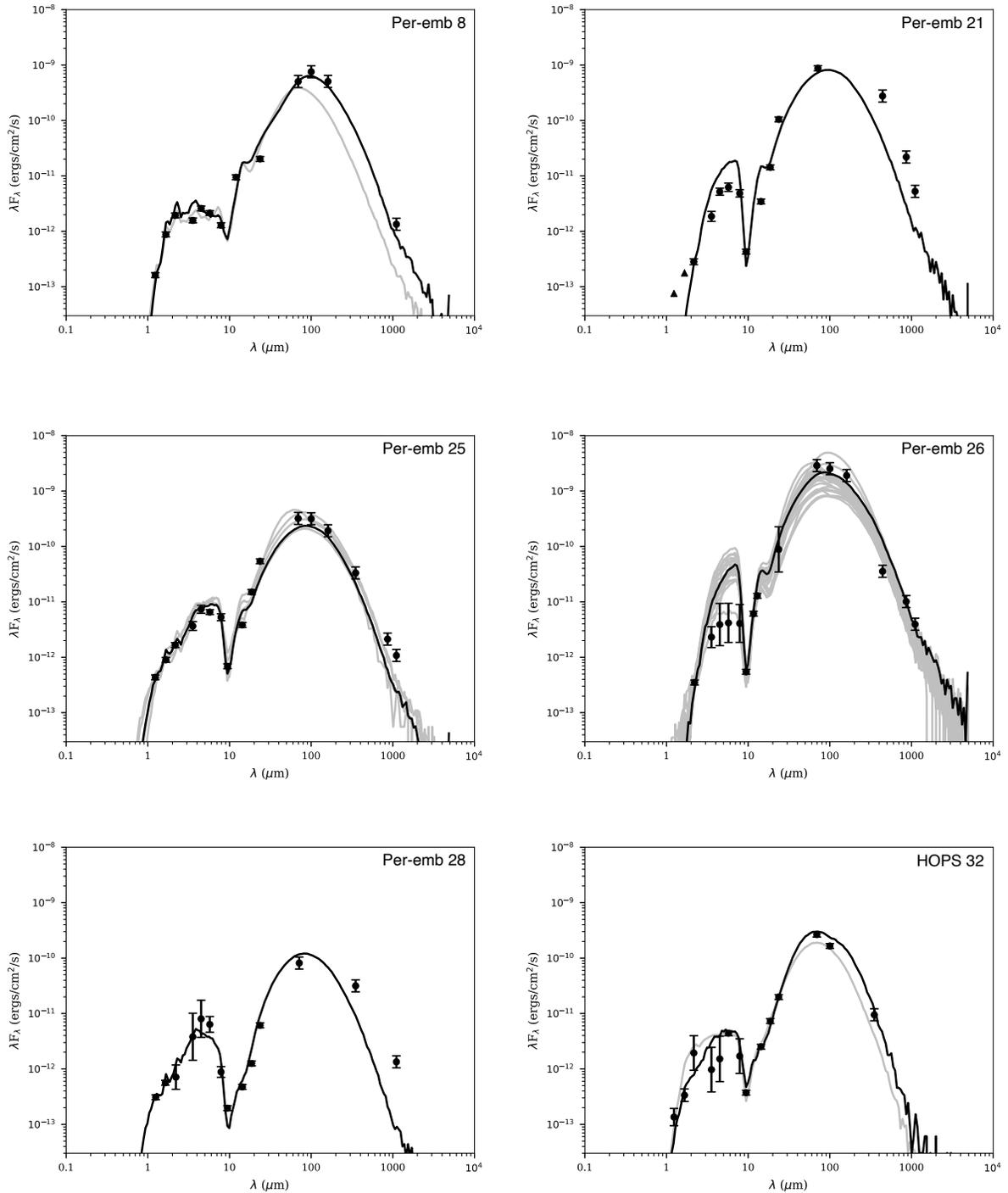}
    \caption{The results of SED fitting for our Class 0 sources using the SED models of \citet{2006ApJS..167..256R}. We show both the best-fit model (black) and the family of models for which $\chi^{2} - \chi^{2}_{best}$ per data point $< 3$ (grey). We detail our methodology and interpret these results in Appendix Section~\ref{sec:SED_fitting}.}
    \label{fig:SEDfits}
\end{figure*}

\begin{figure*}[h]
    \centering
    \includegraphics[width=\textwidth]{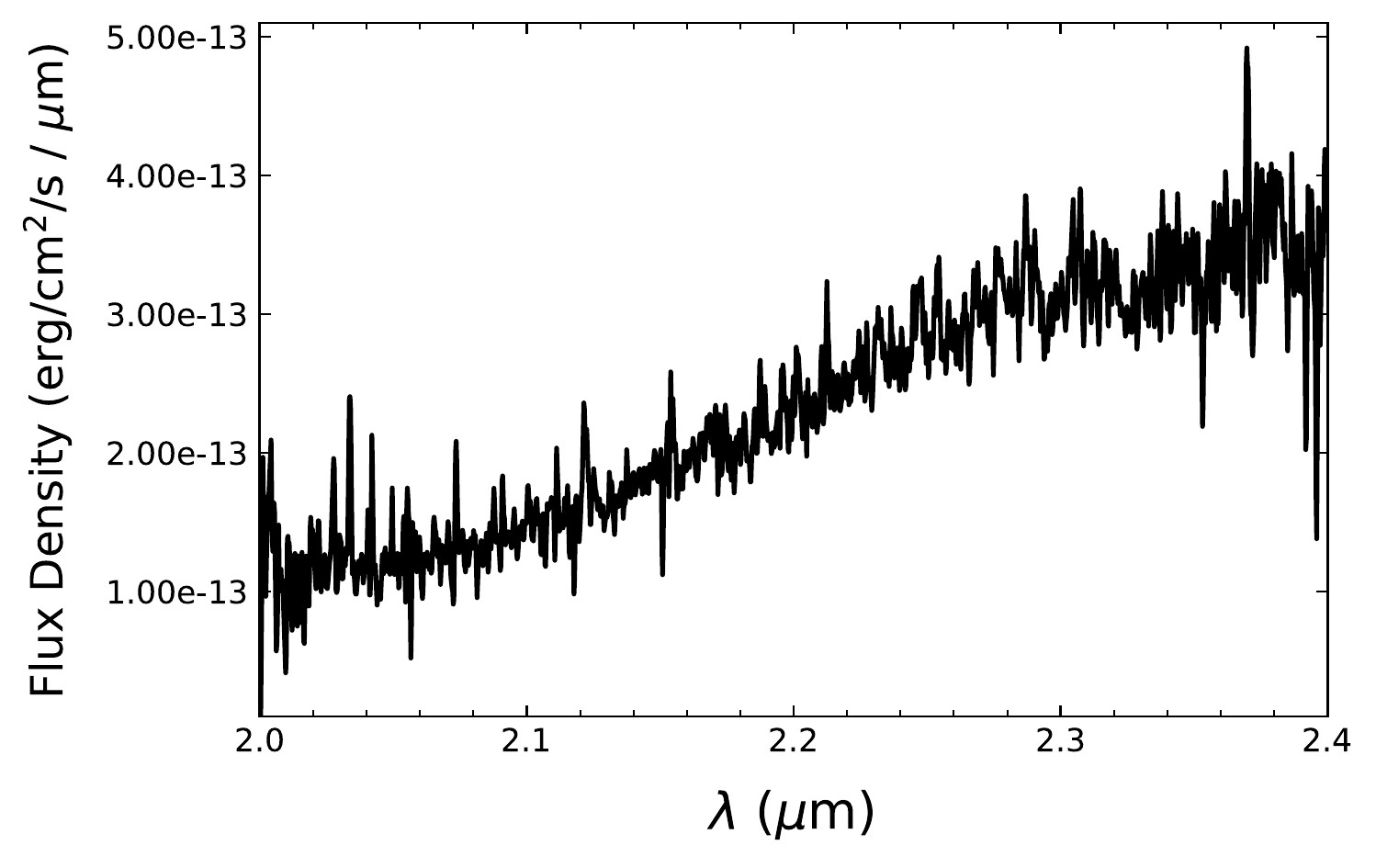}
    \caption{Near-IR MOSFIRE spectrum of HOPS 44 taken on UT Oct 13 for a total integration time of 12 minutes (smoothed by a Gaussian filter for a clearer display). We exclude HOPS 44 from the final sample of Class 0 we analyze in Section~\ref{sec:analysis} due to the low S/N of its observation (Appendix Section B).}
    \label{fig:HOPS44smoothspec}
\end{figure*}

\clearpage

\begin{table*}
\centering
\caption{Derived Parameters from Best-fit SED Models}
\label{tab:SEDparams}
\begin{tabular}{cccccc}
\hline 
 & A$_{v}$\tablenotemark{a} & i\tablenotemark{b} & $\dot{M}$\tablenotemark{b} & M${_c}$\tablenotemark{b} & M$_{env}$\tablenotemark{b} \\
 & mag & deg & M\textsubscript{\(\odot\)}/Myr & M\textsubscript{\(\odot\)} & M\textsubscript{\(\odot\)} \\
\hline
Per-emb 26 & 18.7 & 32 & 2.10E-04 & 7.40E-01 & 19.7 \\
Per-emb 25 & 9.3 & 32 & 5.90E-05 & 2.50E-01 & 8.20E-01 \\
Per-emb 21 & 36.4 & 32 & 1.40E-04 & 4.70E-01 & 5.50E+00 \\
Per-emb 28 & 4.7 & 49 & 2.80E-05 & 1.90E-01 & 3.10E-01 \\
Per-emb 8 & 20.9 & 18 & 7.80E-05 & 2.00E-01 & 3.90E+00 \\
HOPS 32 & 23.9 & 87 & 3.60E-06 & 2.65E+00 & 2.20E-01 \\
\hline
\end{tabular}
\footnotetext[1]{These very approximate, likely lower, foreground extinction estimates were computed with the goal of de-reddening our line luminosities (see Section~\ref{subsec_luminosities}, Appendix Section A).}
\footnotetext[2]{We note these parameters are estimates with large uncertainties from our SED fitting procedure. In some cases, these values likely differ significantly from the true value for the system. Subsequently, they are not used in the analysis of this work and should be considered with care.}

\end{table*}

\end{document}

%% file: Abstract.tex
\begin{abstract}

We present near-infrared \textit{K}-band spectra for a sample of 7 Class~0 protostars in the Perseus and Orion star-forming regions. %We detect emission features probing their near circumstellar environments but do not detect photospheric absorption features. In particular, our spectra reveal a hot inner disk region and highly excited H$_{2}$ in the form of Brackett $\gamma$, CO bands, and H$_{2}$ line emission, respectively. We compare the properties of these features with that of published actively accreting Class I protostars and find they are often consistent in strength and velocity width for Br $\gamma$ and CO.
We detect Br~$\gamma$, CO overtone, and H$_{2}$ emission, features that probe the near circumstellar environment of the protostar and reveal evidence of magnetospheric accretion, a hot inner disk atmosphere, and outflows, respectively. Comparing the properties of these features with those of Class~I sources from the literature, we find that their Br $\gamma$ and CO emission are generally consistent in strength and velocity width. 
The Br~$\gamma$ line profiles are broad and centrally peaked, with FWHMs of $\sim$200 km~s$^{-1}$ and wings extending to $\sim$300 km~s$^{-1}$. The line ratios of our H$_{2}$ emission features, which are spatially extended for some sources, are consistent with shock excitation and indicate the presence of strong jets or a disk wind. 
%Our comparison of Brackett $\gamma$ line profiles finds they are similarly centrally peaked with FWHMs of $\sim$200 km/s and wings extending to $\sim$300 km/s, providing evidence for ongoing magnetospheric accretion. The line ratios of our observed, in some cases extended, H$_{2}$ emission features are consistent with shock excitation and likely from the presence of strong jets or a disk wind. 
Within our small sample, the frequency of CO band emission ($\sim$67\%) is high relative to that of Class~I samples ($\sim$15\%), indicating that Class~0s have high inner disk accretion rates, similar to those of the most actively accreting Class~I sources. Collectively, our results suggest that Class~0 sources have similar accretion mechanisms to the more evolved classes, with strong organized stellar magnetic fields established at the earliest observable stage of evolution. %sharing similar evolutionary characteristics to the more evolved classes, with evidence of inner disk accretion properties similar to actively accreting Class Is and organized magnetic fields being established at even earlier evolutionary stages than previously known.

\end{abstract}

%% file: Introduction.tex
\section{Introduction}\label{sec:intro}

Numerous strides have been made over the past decades to understand the mass accretion of young stellar objects (YSOs), both from an observational and modeling perspective. By the ages of $\sim$1--10~Myr, YSOs fall into the class of T Tauri stars (TTSs, Class II/III). Their dissipated envelopes have allowed for detailed study in both the near-infrared (NIR) and visible. Overall, these phases mark the end of the mass accretion phase, with only the most active Class~IIs exhibiting ongoing accretion.

At the age of $\sim$0.5~Myr, the earlier stage of Class~I protostars are believed to be actively accreting, in the process of building toward their final stellar mass from the reservoir of material in their surrounding envelopes (\citealt{2014prpl.conf..195D}). Sensitive array (SMA, ALMA, NOEMA) observations in the millimeter regime have detected these massive envelopes (\citealt{2009A&A...507..861J}) and used the molecular line emission from their circumstellar disks to place dynamical constraints on the masses of these protostars (\citealt{2014A&A...562A..77H}, \citealt{2017ApJ...834..178Y}, \citealt{2017ApJ...849...56A}, \citealt{2020A&A...635A..15M}, \citealt{2020ApJ...905..162T}, \citealt{2021ApJ...907L..10R}). Additionally, near-infrared observations have given us further insight into the embedded central stars, in some cases detecting photospheric absorption features and emission features (\citealt{2005AJ....130.1145D}, \citealt{2010AJ....140.1214C}) that confirm active accretion.
%Overall, the central objects of these sources are typically in the process of building toward their final stellar mass, accreting the remaining mass from their surrounding envelopes. 

%At the age of $\sim$0.5Myr, the earlier stage of Class I protostars emit the majority of their emission in the mid-to-far infrared (\citealt{2014prpl.conf..195D}). Observations in the millimeter regime have detected the presence of surrounding envelopes and circumstellar disks (\citealt{2009A&A...507..861J}). Additionally, near-infrared observations have given us insight into the embedded central stars, in some cases detecting photospheric absorption features and emission features (\citealt{2005AJ....130.1145D}, \citealt{2010AJ....140.1214C}) that indicate active accretion. Overall, the central objects of these sources are typically in the process of building toward their final stellar mass, accreting the remaining mass from their surrounding envelopes. %Their well-studied characteristics include circumstellar disks of varying sizes, active magnetic fields, chromospheric activity, and winds.

In the past decades, the advent of sensitive, all-sky mid-infrared observatories (\textit{IRAS}, \textit{Spitzer}, \textit{Herschel}) has revealed over a hundred systems younger than Class~Is, designated as Class~0 protostars, in nearby dark clouds. Mounting evidence has found these Class~0s have similar system morphologies to that of the more evolved classes at large scales. Class~0s also exhibit massive surrounding envelopes and energetic, high-velocity outflows (\citealt{1993ApJ...406..122A}), likely powered by strong mass accretion. Recent millimeter observations have begun to reveal the presence of small (on average $\sim$40~AU) circumstellar disks around a fraction of Class~0 objects (\citealt{2013ApJ...779...93T}, \citealt{2018ApJ...866..161S}, \citealt{2019A&A...621A..76M}, \citealt{2020ApJ...890..130T}).

These Class~0 protostars are also thought to be actively accreting, with the majority of their final stellar mass still in the surrounding envelope. Statistical analyses of large YSO samples support a relatively short duration for the Class~0 phase, with lifetimes of $\sim$0.1--0.2~Myr (\citealt{2015ApJS..220...11D}). This brief evolutionary stage likely corresponds to the highest level of overall stellar mass accretion in these objects. Growing evidence finds this accretion is episodic to a degree, with outflow clump structure (\citealt{2001ApJ...554..132A}, \citealt{2015Natur.527...70P}) and CO$_{2}$ ice features (\citealt{2014prpl.conf..195D}) documenting outburst history as frequent and variable. 

%These energetic flows carve their way out of the dense shell of surrounding material, often resulting in bipolar, fan-shaped nebulosities. Likely driven by accretion, strong jets with high velocities ($>$100 km/s) have been observed in the inner regions of these outflows. It has been suspected a toroidal field produced by wound up magnetic field lines near the stellar poles help collimate the inner flow into jets (\citealt{2016ARA&A..54..135H}, \citealt{2016ARA&A..54..491B}). These jets have the ability to shock heat ambient material to hard X-ray temperatures (T $\sim$ 10$^{6}$K). Commonly observed in Class Is and IIs, magnetic field driven X-ray activity has been difficult to confirm for Class 0s (\citealt{2001ApJ...557..747I}, \citealt{2007A&A...463..275G}, \citealt{2009A&ARv..17..309G}). Their low X-ray detection rates are likely a consequence of the high visual extinctions (A$_{v}$ $\geq$ 70) expected from the massive envelopes of the Class 0 phase.

%Growing evidence suspects the driving force of these outflows are strong, collimated jets, as a consequence of shock-heated H$_{2}$ gas in an accretion column. 

Historically, however, high visual extinctions (A$_{V}$ $\geq$ 70) have made detailed study of the Class~0 central stars difficult. High sensitivity observations are required to analyze the, often scattered, near-IR light that leaks through less dense parts of the envelope and the outflow cavities. As a result, little is known about the underlying mechanisms governing young stellar mass accretion and the extent to which they differ between the evolutionary classes. In particular, it is still unclear how nascent protostars manage their angular momentum budget during their formation. 

As they accrete increasingly high angular momentum material from their molecular envelopes, protostars can eventually be spun up to breakup rotational velocities in the absence of an efficient angular momentum loss mechanism. For example, in the idealized case of infall from a singular isothermal sphere with molecular core angular rotation rates of a few times 10$^{-14}$~s$^{-1}$, breakup velocities are reached at a small fraction of a solar mass, $\sim$~0.2\,M\textsubscript{\(\odot\)}(\citealt{1989ApJ...345..959D}, \citealt{2012Natur.492...83T}). This mass regime likely corresponds to the Class~0 or early Class~I phase for a final star of moderate mass ($\sim$0.5--1\,M\textsubscript{\(\odot\)}) given typical accretion rates and lifetimes of the different protostellar stages (e.g. the estimated Class 0 lifetime of $\sim$0.2 Myr with an accretion rate of $\sim 10^{-6}$\,M\textsubscript{\(\odot\)}/yr).
%Growing to Sun-like final masses requires an accompanying angular momentum loss mechanism.

One proposed mechanism to accrete past this expected mass limit is by
%hypothesized the launching of
launching 
a powerful stellar wind that removes angular momentum and drives molecular outflows (\citealt{1988ApJ...328L..19S}). This ``X-wind" picture was later generalized to account for the stellar accretion and wind generation of T Tauri stars, which have 
%been observed to have 
stellar rotational speeds much lower than breakup 
(\citealt{1994ApJ...429..781S}). 
In this case, 
mass accreting from the inner disk is channeled through a stellar magnetosphere onto the star (\citealt{1994ApJ...426..669H}), while angular momentum is fed back to the disk and out into 
%magnetospheric-channeled accretion, in which a strong magnetic field coupled to the circumstellar disk, occurs (\citealt{1994ApJ...426..669H}), creating 
a strong stellar wind that eventually collimates into a jet (\citealt{1995ApJ...455L.155S}). 

Overall, this generalized picture is successful in accounting for many observed properties of T Tauri stars. Study of the younger Class~Is finds they similarly exhibit signs of magnetospheric accretion (\citealt{2005AJ....130.1145D}), with evidence for kilo-gauss strength magnetic fields (\citealt{2009ApJ...700.1440J}). This mounting evidence 
%has supported 
suggests 
that Class~I protostars can also 
%plausibly 
be magnetically coupled to their accretion disks, 
%estimated to be at likely 
although likely at smaller radii 
than Class II sources
(\citealt{2002AJ....124.2185G}) given the observed higher rotation rates for Class~Is than Class~IIs in the same region (\citealt{2005AJ....129.2765C}). It is yet to be seen, however, whether organized stellar magnetic fields can arise in the earlier Class~0 phase, and subsequently whether 
%their accretion is channeled 
they accrete 
directly from their disks 
(e.g., via a boundary layer) 
or via a stellar magnetosphere. 

% It also remains unclear how this accretion evolves temporally as stars transition from their earliest to latest stages. %The accretion-powered luminosities for Class 0 and I objects cannot be reproduced with models assuming a steady, ongoing level of accretion (\citealt{1990AJ.....99..869K}, \citealt{2009ApJ...692..973E}). 
% Protostellar evolution models that assume a steady, ongoing level of accretion overestimate the bulk of observed Class 0 and I luminosities (\citealt{1990AJ.....99..869K}, \citealt{2009ApJ...692..973E}). Instead, models including variable accretion rates that decrease over time (\citealt{2012ApJ...756..118B}, \citeyear{2017A&A...597A..19B}) along with episodic accretion events (\citealt{2011ApJ...736...53O}, \citealt{2012ApJ...747...52D}) have been more successful. \textcolor{red}{Stefan: add observational evidence as outlined in PPVI Dunham talk?} %Given their expected short duration (on the order of years), however, these outbursts are rare, making their statistical study difficult.

We investigate these open questions with new near-IR spectroscopic observations, which probe the near stellar environment of our Class~0s. This wavelength regime offers three primary spectral features to diagnose potential accretion activity. % and characterize Class 0s in the context of the "X-wind" picture. 
Br $\gamma$ emission has been associated with accretion and a signature for magnetospheric activity in Class Is and IIs (\citealt{1996ApJ...456..292N}, \citealt{1998ApJ...492..743M}). CO 
%band 
overtone emission is another strong 
%implication 
indicator
of active accretion, implying warm ($\geq$2000K) temperatures and high ($>$10$^{10}$cm$^{-3}$) local densities (\citealt{1987ApJ...312..297G}, \citealt{1989ApJ...345..522C}). High resolution spectroscopy of these features have strongly argued in favor of a Keplerian, rotating disk as the emitting source (\citealt{1993AAS...18111605C}, \citealt{1996ApJ...462..919N}). Finally, H$_{2}$ emission is often interpreted as a signal for strong, bipolar jets or a wind %expected from evolving YSOs, offering 
and can offer 
insight into Class~0 outflow activity.  

%In addition to Brackett $\gamma$ emission, CO band overtone emission in the near-IR spectra of YSOs has also been interpreted as a strong implication of active accretion. Detected in the low-resolution spectroscopy of embedded protostars and sources with energetic outflows (\citealt{1987ApJ...312..297G}, \citealt{1989ApJ...345..522C}, could cite more?), this emission implies an environment with warm temperatures ($\geq$2000K) and high local densities ($>$10$^{10}$cm$^{-3}$). High resolution spectroscopy of these features have strongly argued in favor of a Keplerian, rotating disk as the emitting source (\citealt{1993AAS...18111605C}, \citealt{1996ApJ...462..919N}). The presence of CO emission offers a rare look into the interaction between the accreting disk and star and has been observed in a small fraction of Class 1s (\citealt{2010AJ....140.1214C}).

%While it is expected the central objects of our Class 0s are undergoing significant mass accretion, the details of this complex process remain unclear. Recent efforts, however, have begun to shed light on the temporal aspect of this evolution. The accretion-powered luminosities for Class 0 and I objects cannot be reproduced with models assuming a steady, ongoing level of accretion (\citealt{1990AJ.....99..869K}, \citealt{2009ApJ...692..973E}). Instead, models including variable accretion rates that decrease over time (\citealt{2012ApJ...756..118B}, \citeyear{2017A&A...597A..19B}) along with episodic accretion events (\citealt{2011ApJ...736...53O}, \citealt{2012ApJ...747...52D}) have been more successful. 
\begin{table*}
\centering
\caption{Journal of Observations}
\label{tab:journalobs}
\begin{tabular}{cccccccc}
\hline
Object & Region & near-IR R.A. (J2000) & near-IR Dec. (J2000) & UT Date & Int. Time & Spatial Extent\tablenotemark{a} & v$_{systemic}$\tablenotemark{b} \\
 &  & (hh mm ss.s) & (\textdegree ' ") &  & (minutes) & (") & (km/s) \\
\hline
Per-emb 26 & Per & 03 25 38.8 & 30 44 06.2 & 2019 Oct 14 & 24 & 3.4 & 5.4 \\
Per-emb 25 & Per & 03 26 37.4 & 30 15 28.4 & 2019 Oct 12 & 32 & 0.9 & 5.8 \\
 &  &  &  & 2019 Oct 13 & 34 & 1 &  \\
Per-emb 21 & Per & 03 29 10.7 & 31 18 20.6 & 2019 Oct 14 & 16 &  &  \\
Per-emb 28 & Per & 03 43 51.0 & 32 03 08.1 & 2019 Oct 14 & 14 & 2.7 & 8.6 \\
Per-emb 8 & Per & 03 44 44.0 & 32 01 36.2 & 2019 Oct 13 & 26 & 0.9 & 11 \\
 &  &  &  & 2019 Oct 14 & 20 & 0.9 &  \\
HOPS 32 & Ori & 05 34 35.4 & -05 39 59.0 & 2019 Oct 12 & 20 & 0.7 & 10 \\
HOPS 44 & Ori & 05 35 10.6 & -05 35 06.3 & 2019 Oct 13 & 12 &  &  \\
\hline
\end{tabular}
\footnotetext[1]{These extents serve as our extractions widths, corresponding to the spatial width of the individual protostellar continnua observed in our 2d spectral images. The low SNRs of our Per-emb 21 and HOPS 44 observations precluded the measurement of their spatial extents.}
\footnotetext[2]{System velocities used to shift our continuum-detected Class~0 spectra (Figure~\ref{fig:compositespec}) into their corresponding systemic rest frames (Section~\ref{subsec_lineprofiles}). For our Perseus sources, we use the derived velocities from the C$^{18}$O(2-1) line fitting in \citealt{2019ApJS..245...21S}. For HOPS~32, we average similarly derived velocity values for other HOPS protostars in \citealt{2020A&A...642A.137N} as a rough proxy.}
\end{table*}

%HOPS 32 was first identified as an Orion (d = 400pc, \citealt{2017ApJ...834..142K}) YSO with mid-infrared excess from \textit{Spitzer} IRAC (3.6, 4.5, 5.8, 8$\mu$m) and MIPS (24$\mu$m) observations \citep{2012AJ....144..192M}.\ Originally classified as a protostellar candidate, addtional observations in the mid- (IRS; cite) and far-IR (\citealt{2010A&A...518L.102A}, \citealt{2013ApJ...767...36S}) helped solidify its Class 0 status via its spectral energy distribtuion modeling by \citealt{2016ApJS..224....5F}. Their best-fit model finds a slightly inclined ($i = 70^{\circ}$) disk with minimal extinction ($A_{v} = 7.7$) relative to other Class 0 sources. 

%Still accumulating its final mass, HOPS 32 has shown evidence of variability, likely due to an accretion outburst. The Young Stellar Object Variability (YSOVAR, \citealt{2011ApJ...733...50M}) survey found HOPS 32 (identified there as ISOY J053435-053959) to oscillate in brightness by $\sim$1mag at both 3.6$\mu$m and 4.5$\mu$m over the course of its 40 day monitoring. As is commonly found with other Class 0 sources, the 0.87mm and 9mm imaging recently conducted by the VANDAM survey \citep{2020ApJ...890..130T} also found a nearby (0.41\arcsec, 166AU), compact companion to HOPS 32. 

In this paper, we analyze our observed near-IR spectra of Class~0 protostars using multiple line diagnostics and investigate potential differences to the published results of more evolved Class I and II sources. In Section~\ref{sec:obs_data}, we describe the target selection criteria used to establish our final Class~0 sample. We also report the source and nature of our data along with our data reduction and processing procedures. In Section~\ref{sec:analysis}, we report the results from the multiple line diagnostics of our near-IR spectra and compare our derived values to that of the literature for Class I and II sources. In Section~\ref{sec:discussion}, we interpret our results in the context of emission line excitation mechanisms and the circumstellar environments of Class~0 sources. Finally, in Section~\ref{sec:sum_conclusion}, we conclude with a brief summary of our findings.

%% file: Obs_data_reduction.tex
\section{Sample and Observations}\label{sec:obs_data}

\subsection{Target Selection}\label{subsec_targetselect}

In our efforts to investigate the accretion and jet processes of Class~0 protostars, we compile a sample of the most promising candidates for observation from Earth's Northern Hemisphere. In particular, we adopt criteria to ensure we have bona fide protostars that likely have yet to accrete the majority of their masses and are bright enough in the near-IR to allow for spectrscopic study. We consider both the Perseus and Orion star-forming regions to maximize the potential for our sample to have numerous candidates capable of spectroscopic study.
%We consider both the Perseus and Orion star-forming regions to probe %potential differences between these low and high mass regions, respectively. 

Protostellar classes are defined observationally by bolometric temperature (T$_{\rm bol}$), or the temperature of a blackbody having the same mean frequency as the observed continuum spectrum. To this end, we began with the lists of suspected Class 0 (T$_{\rm bol} <$ 70~K) protostars in Perseus (c2d \textit{Spitzer} legacy project, \citealt{2009ApJS..181..321E}) and Orion (\textit{Herschel} Orion Protostar Survey, \citealt{2016ApJS..224....5F}). We required significant \textit{Spitzer} IRAC \citep{2004ApJS..154...10F} 3.6--5.8$\mu$m fluxes ($\geq1$ mJy) and inspected mm-interferometer observations (\citealt{2009A&A...507..861J}, \citealt{2010A&A...512A..40M}) when available to verify the existence of massive, extended circumstellar envelopes. For the remaining 
%subsample, (implies that you selected them, but you haven't)
sources,
images from the UKIRT Infrared Deep Sky Survey (UKIDSS, \citealt{2007MNRAS.379.1599L}) were used to measure positions, integrate K-band fluxes, and estimate source FWHM sizes to assess their observation potential. Ideal candidates had compact spatial extent (FWHM $\leq$ 2\arcsec) and were bright enough (K $\sim$ 15--16) to maximize the expected S/N of their observation. %while minimizing the likelihood of continuum veiling of the desired late-type spectral features. 

From this %final 
sample, we observed seven unique targets (5 in Perseus and 2 in Orion). 
%Given our low number of Orion sources, we were unable to concretely investigate differences as a consequence of their different environments. We instead analyze their near-IR Class 0 features collectively.   %\textcolor{red}{Stefan: Describe sources individually in more detail?} 
The object positions, UT dates, and total integration times of our observations are given in Table~\ref{tab:journalobs}.

\subsection{Observations and Data Reduction}

We obtained moderate resolution and moderate S/N near-IR \textit{K}-band spectra of our Class~0 protostars on 2019 Oct 12, 13, and 14 UT with mostly clear skies and 0.6\arcsec\ -- 0.7\arcsec\ seeing. All observations were made with the Keck~I telescope on Maunakea Hawaii, using the MOSFIRE facility spectrograph (\citealt{10.1117/12.856715}, \citeyear{article}) in its long-slit mode. 

Spectra were acquired with a 1\farcs5 (8-pixel) wide slit. 
We measured a spectroscopic resolving power of ${R\equiv\lambda}/{\delta\lambda=}$ 2,400 (120 km~s$^{-1}$) for seeing-limited point sources at a wavelength of 2.24 microns. Some of our objects, however, are fairly extended (their observed spatial extents are listed in Table~\ref{tab:journalobs}), likely reducing our spectroscopic resolving power to R$\sim$1700 in these cases. The plate scale was 0.1798”/pix along the $46\arcsec$ slit length. The order-sorting MOSFIRE K filter was used to record the $\lambda =$ 2--2.4~$\mu$m wavelength range in each exposure. 

Data were acquired in AB pairs of 120~s long exposures, with the telescope nodded 
%10" 
10\arcsec\
along the slit between integrations. The A0 dwarfs HIP~17971 and HIP~27089 were observed to correct the telluric features in the protostar spectra. 

All data were reduced using Pypeit \citep{2020arXiv200506505P}, an open-source Python based data reduction pipeline supporting multiple optical and near-IR spectrographs.\  %\textcolor{red}{Stefan: Is this all accurate? Need to re-check Pypeit documentation}
In Pypeit, a dome flat exposure is first
used to auto-identify and trace the slit edges. Spectra are wavelength calibrated with low-order fits to the OH sky lines present in the science frames. Overall, lower RMS fits were achieved using these lines when compared to fits using the arc spectra of our observed Ne and Ar lamp exposures. Both object and standard images are processed by a cosmic ray masking routine, flat fielded and sky subtracted. Constant extraction widths as a function of wavelength (reported in Table~\ref{tab:journalobs}) are computed from the flux profile of the individual continuua, which do not consider further extended emission. Extracted individual spectra for each pair are then co-added.
%then oriented, processed by a cosmic ray masking routine, and divided by a normalized master flat, generated from our dome flat exposures  %\textcolor{red}{Stefan: bad MOSFIRE pixel treatment?}.For each AB pair, global sky subtraction is performed on the 2-D images via difference imaging and a B-spline fitting procedure. 
To remove instrumental and atmospheric
features, the co-added object spectra were then divided by the corresponding (co-added) spectra of the A0 dwarf observed at similar airmass. When appropriate, we 
%shift 
shifted 
our protostar spectra slightly in wavelength relative to their telluric spectra to minimize telluric artifacts. We 
%multiply 
multiplied 
this result by the (R = 10000 intrinsically, binned to the dispersion of our data) PHOENIX \citep{2013A&A...553A...6H} spectrum of the dwarf stellar model that most closely 
%matches 
matched 
its corresponding standard to achieve our telluric-corrected spectra. Lastly, we flux 
%calibrate 
calibrated 
our spectra by scaling to the K-band fluxes of the corresponding telluric star.

%% file: Analysis.tex
\section{Results}\label{sec:analysis}

We show our near-infrared spectra %the near-IR spectra of our sample (avoids repeating "sample" in next sentence)
in Figure~\ref{fig:compositespec}. 
%\textcolor{red}{What does "absolute look" mean?}
From our sample of seven unique targets, we exclude the 
%low S/N 
spectrum of HOPS 44 from the analysis in the following sections,
because of its low signal-to-noise.
%We share it's 
Its reduced spectrum is shown in Appendix Section~\ref{sec:HOPS44}. We also do not detect the continuum of Per-emb~21
%, only analyzing 
and only analyze here 
its observed H$_{2}$ emission lines.  

Overall, the three main groups of spectral features seen across our near-IR spectra include CO band, \ion{H}{1} Brackett (Br) $\gamma$, and H$_2$ emission, 
%all in emission 
with varying intensity. We do not detect any photospheric absorption features for our sample. In Table~\ref{tab:EWtable}, we report measured equivalent widths (EW) for the emission
%line (CO is a band)
features seen in our sample. All derived strength values presented in Table~\ref{tab:EWtable} correspond to emission spatially coincident with each object’s continuum source. Subsequently, these strength values do not account for emission extended spatially past the extracted continuum extents listed in Table~\ref{tab:journalobs}. For undetected lines, we report 3$\sigma$ upper limits. In addition to EWs, we compute numerous line diagnostics and compare our values to that of the more evolved Class~Is in the following sections.

\subsection{CO Overtone Emission}\label{subsec_COmodeling}

%\textcolor{red}{\sout{Perhaps start here with brief description of the detection -- detect 2-0 and 3-1 bandheads. Does it look similar to CO bandhead emission previously reported from Class Is and energetic Class IIs (e.g., Carr et al. 1993, Najita et al. 1995, and recent refs)? In these older sources, the emission is well explained as arising from the heated atmosphere of the inner disk (e.g., Calvet et al. 1990?). Because disks have low gravity atmospheres similar to giants...}}

We detect emission bands near 2.3 $\mu$m in four of our Class~0 objects (Per-emb~25, Per-emb~26, Per-emb~28, and HOPS~32). Their structure is consistent with CO overtone emission. For a direct comparison, we normalize the bands by fitting the shape of the continuum blueward of the first bandhead, with either a linear fit or a low order quadratic, and dividing them by this estimate.
%In Figure~\ref{fig:compositeCO}, we show the observed continuum-normalized 2-0 and 3-1 CO band profiles, which have also been shifted to the systemic rest frame (Section~\ref{subsec_lineprofiles}). %Similar band shapes are found between our Class 0s, exhibiting a broad blue shoulder followed by the individual overtone lines convolved from our moderate resolution. 
%To assess the nature of our observed CO bandhead emission, we consider the spectra of cool M star giants, which exhibit broad CO absorption features in a low gravity environment similar to this disk atmosphere. We refer to the family of higher-resolution (R=10000) synthetic spectra in the PHOENIX \citep{2013A&A...553A...6H} library to compare to our observed CO bandhead emission. As a starting point, we choose the synthetic spectrum corresponding to typical M giant values (T$_\mathrm{eff}$ = 3500K, log g = 1.5, Fe/H = 0). For an appropriate comparison to our data, we apply the following steps:

To show that these observed features are consistent with spectrally resolved CO emission, we construct a simple emission model using the family of higher-resolution (R=10000) synthetic spectra in the PHOENIX \citep{2013A&A...553A...6H} library. The spectra of cool M star giants exhibit broad CO absorption features in a low gravity environment similar to a disk atmosphere. With our starting PHOENIX spectra (T$_\mathrm{eff}$ = 3500~K, $\log g$ = 1.5, [Fe/H] = 0), we apply the following steps to compare to our data.

The synthetic spectrum is first binned down to match the lower spectral resolution of our data. To invert its CO absorption features into emission, we multiply this spectrum by -1 and then fit its continuum as described for our science objects. We divide the corresponding spectrum by this fit and then add back 2 stellar continnua (setting the baseline to +1). We also allow for an extra scaling factor, equivalent to the potential contribution of veiling, to improve our matches to the individual band amplitudes.
%; we apply an additive factor of 2.5 continuua to its total continuum before division. 
The inverted synthetic spectra are then broadened with a Gaussian filter to match the broader width of each protostar. We manually iterate the width of our Gaussian filter and repeat this process with different PHOENIX model temperatures to achieve the best fit to our data visually (final widths and temperatures shown in Figure~\ref{fig:COmodel_samplefits}). %We iterate the width of the Gaussian filter until a reasonable visual fit to the data can no longer be achieved.\
The reasonable match between the continuum-normalized CO bands in our Class 0s and their corresponding model spectra (Figure~\ref{fig:COmodel_samplefits}) demonstrate that spectrally broadened CO emission is detected. For our observed v=2-0 CO bands, we estimate their velocity width by converting our derived sigma of the Gaussian filter to full width half maxes (FWHMs). We report FWHMs ranging from 200 to 350 km~s$^{-1}$ (Table~\ref{tab:Analysistable}). From our iterative process, we estimate reasonable visual fits allow for a range of $\pm$0.5 sigma units around the best fit sigma value, corresponding to an overall FWHM uncertainty of $\sim$50 km~s$^{-1}$ (except in the case of Per-emb~26, which is much less constrained due to the much lower S/N observation).

We find our observed profiles (black, Figure~\ref{fig:COmodel_samplefits}) are consistent with those observed in more evolved Class Is and energetic Class IIs (\citealt{1993AAS...18111605C}, \citealt{1996ApJ...462..919N}, \citealt{2017MNRAS.465.3039C}). In these older sources, the emission is well explained as arising from the heated atmosphere of the inner disk (\citealt{1991ApJ...380..617C}). Our measured FWHMs are also similar to other previously observed Class~I and IIs (\citealt{1993ApJ...412L..71C}), which have appeared broadened past 100 km~s$^{-1}$. Given the moderate spectral resolution of our data, however, we are unable to constrain other disk properties ($i_{\rm disk}$, $v\sin i$) through a detailed model analysis beyond this estimate.

\subsection{Brackett \texorpdfstring{$\gamma$}{} Equivalent Widths and Line Profiles}\label{subsec_lineprofiles}
%\textcolor{red}{JOAN: perhaps make this section only about Br gamma line profiles, to parallel next section on CO overtone emission.} 
We detect Br~$\gamma$ emission in four of our Class~0 objects (Per-emb~8, Per-emb~25, Per-emb~28, and HOPS~32). We begin by comparing our measured Br~$\gamma$ EWs to that of known Class~Is.  
%In
Figure~\ref{fig:Connelley_CO_Bry}
plots
%, we compare 
the measured EWs of the Br~$\gamma$ line 
%with that of 
and the CO band
%first CO $v=$=2-0 bandhead 
at 2.293~$\mu$m 
for our sources and the Class~Is observed in \citealt{2010AJ....140.1214C}. %We shade the region indicating emission in both features. %and note this subset of Class Is is highly veiled.
%We find Class 0s and Is 
Overall, we find the broad distribution of the observed Br~$\gamma$ and CO v=2-0 band EWs from these Class~Is overlaps with that of our Class~0s.
%Both groups have nominally very similar EWs overall. 
This result implies their near-stellar environments, with accretion and the inner disk sensitively traced by our detected Br~$\gamma$ and CO, appear to be similar between Class~0 and Is.
%\textcolor{red}{Fig 1 seems to show 5 detections? Figure axes need units!}

To examine our observed Br~$\gamma$ line profiles, we first shift them into their systemic rest frame. For our Perseus sources, we use the derived velocities from the C$^{18}$O(2-1) line fitting in \citealt{2019ApJS..245...21S}, which primarily traces the envelopes surrounding our protostars.
%We also inspect the observed line profiles of this emission, which traces the interaction between the hot, inner disk and the protostellar magnetosphere. 
%For our Perseus sources, we first shift the profiles into their systemic rest frame using the derived velocities from the C$^{18}$O(2-1) line fitting in \citealt{2019ApJS..245...21S}. This spectral line primarily traces the envelopes surrounding our protostars. 
%\textcolor{red}{\sout{JOAN: is the C18O tracing the envelope?}}
Although this analysis has not been conducted for HOPS 32, we refer to the similarly derived velocity values for other HOPS protostars in \citealt{2020A&A...642A.137N} and use their average as a rough proxy. Collectively, these shifts are small in magnitude (on the order of $\sim$10 km~s$^{-1}$). 
For a proper comparison, we then normalize our profiles by deriving a linear fit to the continuum redward and blueward of the Br~$\gamma$ emission and dividing the spectrum by this estimate. Using the splot routine in IRAF, we fit each Br $\gamma$ feature with a Gaussian profile to estimate its width. We report the FWHM of the best fit Gaussian in Table~\ref{tab:Analysistable}. All reported FWHMs in Table~\ref{tab:Analysistable} have had our instrumental velocity resolution of 120 km~s$^{-1}$ subtracted out in quadrature.     
We compare these resultant profiles with that of observed Class~Is \citep{2005AJ....130.1145D} in Figure~\ref{fig:Br_gamma_profiles}. Overall, we find the Class~0 line shapes to be very similar to that of the Class~Is 
and Class~IIs, exhibiting a centrally peaked core (FWHM$\sim$200 km~s$^{-1}$) with broad wings
extending to $\pm$300 km~s$^{-1}$. These widths are similar to that found for both T Tauri (\citealt{2001A&A...365...90F}, \citealt{1996ApJ...456..292N}) and Class~I \citep{2005AJ....130.1145D} sources, roughly 100--300 km~s$^{-1}$. We discuss the interpretation of our observed profile structure in Section~\ref{subsec_magnetosphere}. 

\subsection{H\texorpdfstring{$_2$}{} Equivalent Widths and Emission Line Ratios}\label{subsec_ratios}

We detect H$_{2}$ emission lines in six of our Class~0 objects (Per-emb~8, Per-emb~21, Per-emb~25, Per-emb~26, Per-emb~28, and HOPS~32). Among the numerous H$_{2}$ lines observed, we find the 2.1218~$\mu$m H$_{2}$ v=1-0 S(1) emission line is strongest across our sample. Similar to our comparison in Section~\ref{subsec_lineprofiles}, we %compare 
plot the measured EWs of 
the 
Br~$\gamma$ 
%with that of 
and H$_2$ v=1-0 S(1) line between our Class~0s and the Class~Is from \citealt{2010AJ....140.1214C} in Figure~\ref{fig:Connelley_H2_Bry}. We find the lack of an obvious correlation between these quantities unsurprising, given the different suspected locations of their emission source. The detected H$_{2}$ emission likely originates further out in the outflow regions of the system relative to the detected Br $\gamma$ and CO emission, which likely originates near the stellar surface. 
%We find 

Three of our Class~0s (HOPS~32, Per-emb~8, and Per-emb~26) show significantly larger H$_2$ EWs than any of the Class~Is from \citealt{2010AJ....140.1214C}. To quantify the magnitude of this difference, we compute the mean H$_2$ EW between both samples and find they differ by more than the combined standard deviation within each group. We also perform a two-dimensional two-sided KS test between the two samples and find the difference to be statistically significant, reporting a p-value of 0.007. 

We note the spectra of our Class~0s do not show any evidence of photospheric absorption features (Figure~\ref{fig:compositespec}), likely reflective of high veiling in these objects. Comparably, only $\sim$60\% of the Class Is that exhibit both H$_{2}$ and Br~$\gamma$ in emission were found to be highly veiled. The observed Class~Is with low veiling do not appear to exhibit significantly higher H$_{2}$ EWs. %Our Class 0s are likely more highly veiled than this subset but still exhibit moderate to high $H_{2}$ EWs. 
Given this, we may be seeing evidence of higher Class~0 H$_{2}$ flux relative to their photospheric continua in our three outlying Class~0s relative to these observed Class~Is. 

Similar to our Br~$\gamma$ analysis, we fit a Gaussian profile to our observed H$_2$ v=1-0 S(1) emission lines in IRAF and report their FWHMs in Table~\ref{tab:Analysistable} after subtracting out our instrumental resolution (120 km~s$^{-1}$) in quadrature. %In Figure~\ref{fig:H2FWHMcomp}, we find some of Class 0 emission line velocity widths ($\sim$100--150 km s$^{-1}$) to be noticeably broader than that found for the Class Is in \citealt{2010ApJ...725.1100G} ($\sim$10--45 km/s). 
The FWHMs of this unresolved line, however, are also subject to the variable spatial extents of our objects (Table~\ref{tab:journalobs}) and are broadly overestimated. We note our derived values likely probe different parts of the corresponding H$_{2}$ flows depending on the orientation of our objects. %; we do not interpret them in the context of Class I values due to their high uncertainty. 
We also report the velocity shift of the emission centroid relative to the rest wavelength of the line (2.1218~$\mu$m), typically finding velocities of $\geq$ $| 50 |$ km~s$^{-1}$. Overall, the shifted, broad widths of our observed H$_{2}$ emission lines are consistent with collisional excitation in jets or a wind.

%Overall, we see evidence of heavily excited material with signatures of higher strength and velocity H$_{2}$ than observed in the overall population of known Class 1s. %of notably stronger jets in some of our active Class 0s compared to the jets observed in the population of Class 1s. 

The ratios of particular H$_2$ emission lines have the ability to constrain the source excitation mechanism responsible for this observed circumstellar emission. \citet{1995ApJ...446..852G} find the H$_2$ v = 1–0 S(1) (2.1218~$\mu$m) to v = 2–1 S(1) line (2.2477~$\mu$m) ratio is relatively sensitive for differentiating between different excitation mechanisms and calculate predictions for UV excitation, shock-heated gas in the form of a wind, and X-ray excitation of low ionization H$_{2}$. To de-redden our line ratio, we derive rough estimates of the foreground extinction by fitting the spectral energy distributions (SEDs) of our Class~0 protostars. Our methodology is detailed in Appendix Section~\ref{sec:SED_fitting}. We note this correction is generally minimal, given the small difference in wavelength between these individual H$_{2}$ lines. 

We report the reddening-corrected line ratios for our Class~0 sample in Table~\ref{tab:Analysistable}. Similar to the majority of the Class Is studied in \citet{2010ApJ...725.1100G}, we find the values of our Class~0s are most consistent with collisional excitation via shocked gas in a wind or X-ray excitation but not UV excitation (Figure~\ref{fig:H2lineratiocomp}). The more moderate S/N in the fainter H$_{2}$ line transitions precludes our ability to compare this result with other line ratios.

%\textcolor{red}{\sout{JOAN: Fig 11 seems to show that the H2 emission from Class Is spans a broad range, including winds, X-rays, and reaching close to UV even! Class 0s seem to be somewhere between winds and X-rays. Does that mean it's only shocks?}}

% Collisional excitation of H$_2$ via shocked gas also has the potential to reproduce the observed H$_2$ line ratios. \citealt{1989MNRAS.236..409B} predict the v = 1–0 S(1) to v = 2–1 line ratio for varying gas temperatures and densities. We find the closest agreement to our observed ratio comes from a 2000K gas in LTE. Similarly, \citealt{1991ApJ...383..205W} calculate multiple H$_2$ line intensities relative to the v=1-0 S(1) line. Their values for similar gas properties (LTE collisional excitation at T$_{neutral} = 2000$K and T$_e = 3000$K) are closest in agreement to the observed values for all of our sources in the v = 1–0 S(1) to v = 2–1 S(1) line ratio. The more moderate S/N in the fainter H$_{2}$ line transitions, however, precludes our ability to compare with other line ratios. 
\subsection{H\texorpdfstring{$_2$}{} Extended Emission}\label{subsec_extended}

We find HOPS~32, Per-emb~26, and Per-emb~8 (the strongest H$_{2}$ EW sources, Figure~\ref{fig:Connelley_H2_Bry}) also exhibit H$_{2}$ emission extended from the continuum; an example is shown in Figure~\ref{fig:Per8extend}. Our observations of Per-emb~21, while not sensitive enough to detect the continuum, similarly detect variably extended H$_{2}$ emission. Per-emb~25 and Per-emb~28 do not show evidence of any off-axis emission. From our 2d images, we report the measured angular extent of the H$_{2}$ 1-0 S(1) line in Table~\ref{tab:Analysistable}. In all cases, we find the emission is far extended spatially ($\geq$5\arcsec, i.e., $\sim$1250~AU across for our Perseus sources, and 8\arcsec, i.e., $\sim$3200~AU for HOPS~32). 

We reference the literature for known outflows corresponding to our Class~0 sources in an attempt to find other potential tracers of our extended emission. Numerous Perseus protostars have had their outflow features identified via both sensitive molecular line emission tracing (\citealt{2016ApJ...820L...2L}, \citealt{2017ApJ...846...16S}, \citealt{2018ApJ...867...43T}) and H$_{2}$ 2.122-$\mu$m imaging surveys \citep{2008MNRAS.387..954D}. In Table~\ref{tab:PAtable}, we compare published outflow position angles (PAs) with the slit PAs from our observations when possible. 

We find that, as expected, extended emission is present when the slit PA is aligned with the PA of a known CO outflow. As an example, Per-emb~8 was observed in \citet{2018ApJ...867...43T}, which found evidence for strong, rotational CO 2-1 emission northwest of the system at a PA of 313\textdegree\ lacking a southeast component. We find this asymmetry reflected in our data given the close alignment with our slit PA, showing the extended H$_{2}$ emission spatially offaxis from the continuum in only one direction (Figure~\ref{fig:Per8extend}). Alternatively, the observed H$_{2}$ outflow feature identified for Per-emb~25 \citep{2008MNRAS.387..954D} has a PA close to orthogonal with our slit PA. In this case, its lack of extended H$_{2}$ emission is not surprising.

%Across our sample, we find good agreement between their relative axial alignment of outflows and the presence of extended emission. As an example, Per-emb 8 was observed in \citet{2018ApJ...867...43T}, which found evidence for strong, rotational CO 2-1 emission northwest of the system at a PA of 313\textdegree\ lacking a southeast component. We find this asymmetry reflected in our data given the close alignment with our slit PA, showing the extended H$_{2}$ emission spatially offaxis from the continuum in only one direction (Figure~\ref{fig:Per8extend}). Alternatively, the observed H$_{2}$ outflow feature identified for Per-emb 25 \citep{2008MNRAS.387..954D} has a PA close to orthogonal with our slit PA. In this case, the lack of extended H$_{2}$ emission is not surprising. The slit PAs for observations of the rest of our sample only marginally align with detected outflows (except in the case of HOPS 32, which does not have published axial outflow information). However, we find the relative angular extent of the outflow features (e.g. wide in the case of Per-emb 26/21, narrow for Per-emb 28) explains the presence, or lack their of, extended emission in their data, respectively.

\subsection{Line Luminosities}\label{subsec_luminosities}

To compute individual line luminosities, we deredden our measured line fluxes using our derived foreground extinction values (Table~\ref{tab:Analysistable}, Appendix~\ref{sec:SED_fitting}). We note these values are estimates with high uncertainty given the complicated physical nature of our Class~0s (e.g., non-symmetrical envelopes, scattered light). Subsequently, it is difficult to draw robust conclusions from our derived line luminosities. Despite these limitations, we tentatively compare these values with that observed in more evolved systems as a first look between Class~0s and Is. 

%We derive rough estimates of the foreground extinction by fitting the spectral energy distributions (SEDs) of our Class 0 protostars. Our methodology is detailed in Appendix Section~\ref{sec:SED_fitting}. 
%We assume a distance of 250 parsecs for our Perseus sources (\citealt{2006ApJ...638..293E}) and 420 parsecs for our Orion sources (\citealt{2017ApJ...834..142K}).

In Figure~\ref{fig:LogH2lumcomp}, we compare our derived, dereddened H$_{2}$ v=1-0 S(1) line luminosity with that of the Class~Is from \citealt{2010AJ....140.1214C}. We find our Class~0s appear to exhibit a wider range of H$_{2}$ v=1-0 S(1) line luminosities than Class~Is, with a roughly similar median value between both samples (vertical red and black dashed lines, left). Given our foreground extinction estimates are likely lower limits to the true extinction, we consider the effect on this comparison after increasing our extinction estimates to a potentially more appropriate value. We consider the highly accreting Class I protostars in Rho Ophiuchi as a rough proxy, which have continuum brightness of \textit{K}$\sim$10 mag with estimated A$_{v}$ of $\sim$20 mag. An overall foreground extinction estimate of A$_{v}$ $\sim$50 magnitudes would be consistent with our Class 0 continuum brightnesses of \textit{K} $\sim$15-16 mag after scaling to the distance of Perseus. While the true extinctions to our Class 0 sources could be larger than this estimate, it is unlikely to be considerably so (i.e., A$_{v}$ $\geq$ 100) given that we detect emission from the near-stellar region and
our near-IR flux is likely dominated by scattered light. 
%and the success of our observations. 
Although the application of this correction to our median Class 0 H$_{2}$ v=1-0 S(1) line luminosity (red dashed line, right) implies a potential difference between these two samples, we emphasize it remains unclear the extent to which these line luminosities between these two populations differ until more precise Class 0 extinction estimates can be derived.

We also compare our Br~$\gamma$ luminosities with that of observed YSOs, referencing low mass Class~I and IIs from the literature. In Figure~\ref{fig:Brylumcomp}, we show their respective luminosity distributions. %Although it is perhaps surprising that our Class~0s have median Br~$\gamma$ luminosities $\sim$100 times smaller than those of Class~Is, the discrepancy could be resolved if we have systematically underestimated the extinctions of Class~0 systems by a factor of 2. 
Similar to Figure~\ref{fig:LogH2lumcomp}, we show the derived Br $\gamma$ line luminosity median for our Class 0s using the foreground extinction estimates from our SED fitting (vertical red dashed line, left) and a corrected extinction value of 50 mag for all of our Class 0 sources (vertical red dashed line, right). %We note using the latter value appears to resolve the visual discrepancy seen in the former case between the line luminosities of our Class 0s and the low-mass Class Is and IIs of \citealt{2008A&A...479..503A} and \citealt{1998ApJ...492..743M}, respectively.
With an A$_{v}$=50, the median Class 0, I, and II Br$\gamma$ luminosities are similar. More precise Class 0 extinction estimates are needed to make a reliable quantitative comparison.
%However, we reiterate that more precise Class~0 extinction estimates are still required to confirm their overall similarity, or lack their of. 
%\textcolor{red}{JOAN: Alternative wording might be "With an Av=50, the median Class 0, I, and II Br$\gamma$ luminosities are similar. More precise Class 0 extinction estimates are needed to make a reliable quantitative comparison."}

%Overall, it remains unclear the extent to which the line luminosities of these two populations differ until more precise Class~0 extinction estimates can be derived. 
We note our high extinction uncertainties also preclude our ability to derive robust mass accretion rate estimates from our Br~$\gamma$ luminosities, as has been done for Class~Is and IIs (\citealt{1998ApJ...492..743M}, \citealt{2021arXiv210303863F}).

%There is some evidence that our Perseus Class 0s appear to be less luminous in Br $\gamma$ relative to these older samples.

%For both our H$_{2}$ and Br $\gamma$ luminosity comparisons, however, it is difficult to draw robust conclusions without accurate extinction estimates. The visual extinction at the Class 0 phase can be significant (A$_{v}$ $\sim$ 70 magnitudes). Our derived Class 0 extinction values are likely lower limits and may differ greatly from the true visual extinction toward each protostars. The slight visual difference seen in the relative Br $\gamma$ luminosities between our Class 0s and that of Class 1s is resolved by increasing our extinction estimates by a factor of 2. 

%Br $\gamma$ luminosities for Class Is and IIs have been used to estimate their mass accretion rates (\citealt{1998ApJ...492..743M}, \citealt{2021arXiv210303863F}). Despite their apparent lower Br $\gamma$ luminosities, our Class 0s are likely accreting at higher rates than their more evolved counterparts. It is likely our Class 0 Br $\gamma$ luminosities do not reflect their underlying mass accretion rate due to significant extinction. %Understanding the extinction correction is critical if we want to be able to use measured Br gamma emission from Class 0s as a measure of their stellar accretion rate. 

%serves as a clear accretion indicator for Class 0s as compared to Class I/IIs.  

%% file: Discussion.tex
\section{Discussion}\label{sec:discussion}

The line diagnostics analyzed in Section~\ref{sec:analysis} probe the near-circumstellar environment of our Class~0 sample and lend insight into the accretion and outflow processes underway in these YSOs. Here, we compare our measured Class~0 properties with those of well-studied Class~I and II sources to investigate the nature of the near-circumstellar environment at the earliest observable stage of evolution.   

%We characterize the near circumstellar environments of our Class 0 sample with our line diagnostic analysis in Section~\ref{sec:analysis}. These results give insight into the mechanisms surrounding the active accretion and outflows expected in these YSOs, which have yet to be well defined with observational evidence. We interpret these results in the context of well-studied Class I and IIs, finding many similarities with our sample at this even earlier evolutionary phase. 

%\subsection{Compact Magnetosphere Paradigm}\label{subsec_magnetosphere}
\subsection{Class 0 Stellar Magnetospheres}\label{subsec_magnetosphere}

In Section~\ref{subsec_lineprofiles}, we find our Class~0 sample has broad, centrally peaked (Figure~\ref{fig:Br_gamma_profiles}) Br~$\gamma$ line profiles (FWHM $\sim$ 200 km~s$^{-1}$) with wings extending to high velocity (300 km~s$^{-1}$). Similar profiles are observed in many Class~II (\citealt{1996ApJ...456..292N}, \citealt{2001A&A...365...90F}) and Class~I (\citealt{2005AJ....130.1145D}, \citealt{2010AJ....140.1214C}) sources. 
%This finding is very similar to what has been seen in many observed Class II (\citealt{1996ApJ...456..292N}, \citealt{2001A&A...365...90F}) and Class I (\citealt{2005AJ....130.1145D}, \citealt{2010AJ....140.1214C}) sources. 
%\textcolor{red}{\sout{Reference here older work by Muzerolle et al, Najita et al, Folha \& Emerson, and more recent refs on Class IIs and Tom's and other's work on Class Is.}}
%
%The dominant source of Br $\gamma$ emission for YSOs has long been debated (cite Carr 1993, Najita 1996). Proposed physical mechanisms require high velocities to reproduce the observed broad line widths. 
In these more evolved sources, the Br~$\gamma$ profiles are often considered strong evidence for magnetospheric accretion, in which the inner region of the hot circumstellar disk is channeled by stellar magnetic field lines onto the surface of the star at high infall velocities. %The line asymmetries we see in the case of HOPS 32 and Per-emb 28, which show more flux in the blue wing of the Br $\gamma$ feature relative to its centroid (Figure~\ref{fig:Br_gamma_profiles}), may be evidence of a high inclination viewing angle along the accretion column.

Models of emission from magnetospheric accretion in the Balmer lines (\citealt{1992ApJ...386..239C}, \citealt{2012MNRAS.421...63R}) and Br~$\gamma$ (\citealt{1998ApJ...492..743M}) have had success reproducing these observed profile shapes. This interpretation is supported by measurements of strong (approaching 3~kG) stellar magnetic fields in Class~II (\citealt{2007ApJ...664..975J}) and Class~I (\citealt{2009ApJ...700.1440J}) sources. 

While Class~0 magnetic field strengths have yet to be estimated directly, the similarity between our Class~0 Br~$\gamma$ profiles and those of these more evolved sources argues in favor of a shared accretion mechanism. These results provide evidence of protostellar magnetic fields and magnetospheres being established very early in the star formation process, at the earliest times yet probed.

\subsection{Comparison of Class 0 and Class I/II Accretion Activity}

%The presence of both Br $\gamma$ and CO emission in our Class 0 spectra are strongly suggestive of active accretion. The bolometric luminosities of our Perseus Class 0s are high, with a median value greater than that of known Perseus Class 1s (include figure?). Given their lower mass, Class 0s are expected accrete at higher rates than Class 1s to power these high bolometric luminosities. 

Although our Class~0 sample is small, we note the high frequency of our detected spectral features in emission. There have been few other Class~0s that have had their near-IR spectroscopic observations analyzed for both emission and absorption components. Serpens S68N also exhibited H$_{2}$ emission but was absent in Br~$\gamma$ and showed CO bands in absorption rather than emission (\citealt{2018ApJ...862...85G}). We report the incidence rates of these features in emission among our 5 continuum-detected Class~0s (Figure~\ref{fig:compositespec}) and Serpens S68N in Table~\ref{tab:Emissionstats}. We compare these rates with those of Class~Is in nearby molecular clouds (\citealt{2005AJ....130.1145D}). These Class~I statistics are similar to what has also been observed in a larger, all-sky sample of Class~I protostars (\citealt{2010AJ....140.1214C}).

We find similarly high rates of Br~$\gamma$ emission between the Class~0 and Class~I samples, consistent with the active mass assembly expected at these young ages. For both CO and H$_{2}$, however, we find notably higher emission incidence rates for Class~0s relative to Class~Is. %This finding may be an indication of more frequent accretion activity in the Class 0 phase. %Larger samples will be required to confirm whether this behavior seems to also extends to the Class 0 phase.
%\textcolor{red}{(In the last sentence we should instead be comparing the CO incidence rate of Class Is vs. IIs. That would be more direct?)}

The high prevalence of CO emission among our Class~0s suggests differences in the inner accretion disks of these objects relative to that of Class~Is. %This is likely reflective of a higher Class 0 accretion rate in the inner disks, which drives the warm, dense inner disk properties associated with CO emission. 
Our observations find that the atmospheres of Class~0 inner disks frequently achieve the warm ($\geq$2000~K), dense environments ($>$10$^{10}$~cm$^{-3}$) required for CO emission, strongly suggestive of a higher average disk accretion rate for Class~0s than Class~Is. We perform a Fischer's exact test and find this result is statistically significant, with a nominal p-value of 0.01. 
A high accretion rate for Class~0 sources is consistent with the low incidence rate of photospheric absorption features among Class~0s relative to Class~Is. %This result is also supported by the observed low incidence of photospheric absorption features among Class 0s relative to Class Is. 
Currently, absorption features (\ion{Na}{1} and \ion{Ca}{1}, 2.21 and 2.26~$\mu$m respectively) have only been observed in Serpens S68N, where it is detected weakly. The low incidence rate is likely a consequence of higher veiling in Class~0s %as a consequence of their inner warm disks.
produced by strong continuum emission from their inner warm disks.

%The higher prevalence of H$_{2}$ emission among our Class 0s may also be connected to an overall higher level of accretion at this younger stage relative to Class 1s. As noted in Section~\ref{subsec_magnetosphere}, our $H_{2}$ findings are consistent with the predictions of "X-wind" paradigm of YSO evolution. In this picture, a high level of disk accretion (implied by our observed high rate of CO emission) drives active jets and/or a strong disk wind that interacts with the surrounding envelope material.

The high disk accretion rate implied by these results (CO emission, high veiling) could be similar to that inferred for active Class~Is (\citealt{2021arXiv210303863F}) and EXors (\citealt{2010ApJ...719L..50A}), two groups that %have similarly exhibited our observed Class 0 near-IR features in emission
have also shown near-IR features in emission and have estimated accretion rates of 
($\sim$10$^{-7}$--10$^{-6}$ M\textsubscript{\(\odot\)}~yr$^{-1}$). 
%\textbf{We note our highly veiled, near-IR spectra also appear similar to that found by \citealt{2020ApJ...905..162T} for HOPS 370, a suspected intermediate mass protostar ($\sim$2.5M\textsubscript{\(\odot\)}) with an estimated accretion rate of $\sim$10$^{-5}$M\textsubscript{\(\odot\)}~yr$^{-1}$). 
At a higher disk accretion rate, more similar to that of FU Oris ($\sim$10$^{-4}$ M\textsubscript{\(\odot\)}~yr$^{-1}$), the CO overtone bands would appear in absorption (\citealt{1991ApJ...380..617C}) and the magnetosphere would be crushed, reducing or eliminating the Br~$\gamma$ emission. At much lower disk accretion rates typical of Class~II T Tauri stars ($\sim$10$^{-8}$ M\textsubscript{\(\odot\)}~yr$^{-1}$), the inner disk would not be warm enough over a large enough vertical column density to produce CO overtone emission.

%However, our evidence for magnetospheric accretion (Br $\gamma$) also implies an upper limit to the accretion strength. FU-Ori levels of accretion ($\sim$10$^{-4}$ M\textsubscript{\(\odot\)}/yr) are thought to crush the magnetosphere to the stellar surface, explaining the very weak or absent Br $\gamma$ emission (\citealt{2010AJ....140.1214C}) and CO band absorption (\citealt{1991ApJ...380..617C}) observed in these objects. Although currently unclear, the kilogauss strength magnetic fields seen in Class I (\citealt{2009ApJ...700.1440J}) and IIs (\citealt{2007ApJ...664..975J}), could plausibly be established by a $\sim$0.1-0.2M\textsubscript{\(\odot\)} stellar core in the Class 0 phase. 

%These features are not observed together in the near-IR spectra of TTS, which only exhibit Br $\gamma$ emission. This is likely a consequence of the relatively lower level of accretion observed for active TTS (typically $\sim$10$^{-8}$ M\textsubscript{\(\odot\)}/yr), which is not high enough to reproduce the disk temperatures needed for CO bandhead emission (\citealt{1991ApJ...380..617C}).

%Similarly, the majority of Class Is do not exhibit both features in emission. This behavior is not unexpected given the majority of Class Is have likely already accreted the bulk of their stellar mass. The subsample of Class Is that do exhibit both features in emission may be accreting at similar levels as our Class 0s, potentially transitioning between the Class 0 and I stages. 

At a disk accretion rate of $\sim$10$^{-6}$ M\textsubscript{\(\odot\)}~yr$^{-1}$, a protostar can sustain a stellar magnetosphere (and produce the Br~$\gamma$ line profiles we observe) if it has developed a strong, organized stellar magnetic field. If our Class~0 sources have a protostellar mass of 0.2~M\textsubscript{\(\odot\)} (as measured for the similarly low-luminosity Class~0 source L1527 IRS; \citealt{2012Natur.492...83T}) and a stellar surface gravity ($\log g$) of 2.38 (as measured for Serpens S68N; \citealt{2018ApJ...862...85G}), protostellar radii are $\sim$4--5~R\textsubscript{\(\odot\)}. If the Class~0 sources further have a disk accretion of $\sim$10$^{-6}$ M\textsubscript{\(\odot\)}~yr$^{-1}$ and their stellar magnetospheres truncate the inner disk at $\sim$2 stellar radii (as measured for the Class~I YLW~15; \citealt{2002AJ....124.2185G}), the required protostellar magnetic field strength is $\sim$1~kG (\citealt{2016ARA&A..54..135H} and references therein). Thus, the CO overtone emission, strong veiling, and Br~$\gamma$ we observe are all plausibly consistent with Class~0 inner disk accretion rates of $\sim$10$^{-6}$ M\textsubscript{\(\odot\)}~yr$^{-1}$ and approximately kilogauss strength protostellar magnetic fields.

Even within our small sample, several sources depart from the simple picture described above. %With the sample of Class 0s observed in the near-IR growing, we briefly comment on the more unusual systems. Per-emb 26 exhibits CO overtone emission but does not exhibit Br $\gamma$ emission like the majority of our Class 0 sample. 
Unlike the majority of sources in our Class~0 sample, Per-emb~26 exhibits CO overtone emission but no Br~$\gamma$ emission. This object may be in an intermediate state, with a slightly higher accretion rate than the majority of our Class~0s. An accretion level of this magnitude could plausibly suppress the stellar magnetosphere, explaining the absence of Br~$\gamma$, while still remaining below an FU Ori accretion level, allowing for CO emission (rather than absorption). As a variant in a different direction, the recent near-IR spectrum of Serpens S68N lacks Br~$\gamma$ emission and exhibited CO bands in absorption (\citealt{2018ApJ...862...85G}). Serpens S68N is likely in a state of lower disk accretion than our Class 0s, with low veiling indicated by its observed photospheric absorption features (\ion{Na}{1}, \ion{Ca}{1}). Its modest derived $\log g$ value (2.38) is more consistent with a low-gravity stellar photosphere than a disk. Its inflated radius implied by this derivation may be evidence of a significant, recent accretion episode, depleting the reservoir of available accreting material in its disk. % indicating the lack of a very active circumstellar disk due to an extremely young age, as seen in other Class 0 objects (\citealt{2013ApJ...779...93T}). %Overall, these findings point to our Class 0 sample potentially being in a higher accretion rate state relative to that of Serpens S68N. 

\subsection{Nature of the H\texorpdfstring{$_2$}{} Emission}\label{subsec_H2nature}

We interpret the results from our multiple line diagnostics in Section~\ref{sec:analysis} to assess the nature of the H$_{2}$ emission seen in our Class~0 sample. We find our Class~0 H$_{2}$ emission properties are often similar to that of known Class~Is but in some cases enhanced, indicating higher EWs (Figure~\ref{fig:Connelley_H2_Bry}) and large spatial extensions (Figure~\ref{fig:Per8extend}). %We note the prevalence of extended H$_{2}$ emission observed in our sample relative to Class Is may be reflective of the likely stronger accretion expected in Class 0s.% still building towards the bulk of their stellar mass.

In our attempt to constrain the most likely H$_{2}$ excitation, we derived 1–0 S(1)/2–1 S(1) line ratios that argue in favor of either shock-heated gas or X-ray excitation. If excited by collisions in shocks, the molecular hydrogen ortho:para ratio can also be derived with little sensitivity to the H$_2$ rotation temperature from the v=1-0 S(1) and v=1-0 S(0) line ratio \citep{2007A&A...469..561K}. From their equation 5, we find ortho:para ratios ranging from 2.3--2.9 assuming a rotational temperature of 3500~K across our sample. This result provides further evidence of molecular hydrogen emission in shocks as \citet{2000A&A...356.1010W} finds the hydrogen atom exchanges in shocks set the high temperature limit to o/p $\leq$ 3.

We query the literature for observations of X-ray activity in our Class 0 sample, which may also suggest ongoing accretion. X-ray activity can be quantified via an object's hardness ratio [HR = (N$_{h}$ - N$_{s}$)/(N$_{h}$ + N$_{s}$)], where N$_{h}$ and N$_{s}$ are the detected counts in the hard (2--8~keV) and soft (0.5--2 keV) energy bands, respectively. \citet{2004AJ....127.3537S} computed the extinction-corrected hardness ratios between a sample of non-accreting weakly-lined TTS and accreting classical TTS, finding the latter population was more likely to have high HRs. 

Despite the likely high visual extinctions expected towards our Class~0s, we find Per-emb~8 has been detected by \textit{Chandra} measurements (\citealt{2016ApJS..224...40W}). Following the analysis of \citet{2004AJ....127.3537S}, we correct our derived value for extinction. The corrected HR value for Per-emb~8 ($\sim$0.3) lies near the very high end of the distribution for PMS stars. We find this result, while not a definitive indicator for accretion, is consistent with our cumulative evidence that our Class~0s are likely undergoing active accretion. In particular, the strong X-ray activity observed in Per-emb~8 could be further evidence of organized magnetic fields being plausibly established by Class~0s, supported by the recent observations of Class~0 HOPS~383 (\citealt{2020A&A...638L...4G}).

Collectively, these findings are consistent with the properties expected from a strong jet or wind. The recent observation of an extended jet observed toward Per-emb~8 (\citealt{2018ApJS..238...19T}) argues in favor of the former interpretation as a potential source for the observed X-ray emission. We note our derived H$_{2}$ line ratios are consistent with mechanical excitation (Figure~\ref{fig:H2lineratiocomp}), which may be due to the shocked gas expected in a strong protostellar jet.
Overall, the observed high velocity, often extended, H$_{2}$ emission in this work may be evidence of the driving force behind the molecular outflows commonly observed in Class~0s. Future observations of additional Class~0s will be required to further characterize the predominant inner H$_{2}$ emission source at this evolutionary stage.
%e.g., as predicted for ``X-winds" (\citealt{1995ApJ...455L.155S}) of YSO evolution. 
%Along with our magnetospheric findings (Section~\ref{subsec_magnetosphere}), our observed H$_{2}$ emission properties provide an additional component of evidence that this paradigm, consistent with observed Class~I and IIs, could plausibly extend to the earlier Class~0 phase. However, a larger sample of Class~0s will be required to rule out alternative theories. 

%% file: Summary_Conclusion.tex
\section{Summary and Conclusion}\label{sec:sum_conclusion}

We present new observations of near-infrared \textit{K}-band spectroscopy for a sample of 7 Class~0 protostars in the Perseus and Orion molecular clouds. The H$_{2}$, Br~$\gamma$, and CO v=2-0 band features we detect probe the inner circumstellar environments of these systems (Figure~\ref{fig:compositespec}). The lack of observed stellar absorption features indicates high continuum veiling, likely from an inner disk with a high accretion rate. 
%Our spectra do not detect photospheric absorption but instead probe the inner circumstellar environments of these systems, exhibiting H$_{2}$, Br $\gamma$, and CO v=2-0 band emission (Figure~\ref{fig:compositespec}). With multiple line diagnostics, we report the properties of this emission for our Class 0s and compare with that of previously published Class Is. We summarize the conclusions from our findings below.

\begin{enumerate}
\item Our Class~0 Br~$\gamma$ line profiles are broad, centrally peaked, and similar in shape to those of previously observed Class~Is (Figure~\ref{fig:Br_gamma_profiles}) and strongly suggestive of magnetospheric accretion. This result suggests that organized stellar magnetic fields are established very early, at the earliest observable stage of their evolution. 

%Our Br $\gamma$ line profile comparison between our Class 0s and previously observed Class Is (Figure~\ref{fig:Br_gamma_profiles}) finds them to be similarly centrally peaked, to the ability we can discriminate with our limited velocity resolution (120 km/s). In Section~\ref{fig:Br_gamma_profiles}, we argue this line morphology may be evidence of magnetospheric accretion, and thus evidence of an established magnetic field at the even earlier protostellar evolutionary state of Class 0.

%In many cases, the derived emission properties of our Class 0 sample are consistent with that found for Class Is. The measured strengths (Figure~\ref{fig:Connelley_CO_Bry}, Table~\ref{tab:EWtable}) and widths (Figure~\ref{fig:Br_gamma_profiles}, Table~\ref{tab:Analysistable}) of both Br $\gamma$ and CO have similar distributions between the two classes. We find our Class 0 H$_{2}$ emission properties are in some cases enhanced relative to Class Is, indicating higher EWs (Figure~\ref{fig:Connelley_H2_Bry}), velocity widths (Figure~\ref{fig:H2FWHMcomp}), and spatial extension (Figure~\ref{fig:Per8extend}). Similar to the majority of Class Is, our derived Class 0 H$_{2}$ line ratios argue in favor of shocks as their source excitation mechanism. %We also find our Class 0 H$_{2}$ line luminosities to be consistent with that of Class Is. 

\item The high frequency of CO band emission observed in our Class~0s (4/6 $\cong$ 67\%) relative to observed Class~Is (8/52 $\cong$ 15\%) is strongly suggestive of a higher average disk accretion rate for Class~0s. A disk accretion rate of $\sim$10$^{-6}$ M\textsubscript{\(\odot\)}~yr$^{-1}$ is consistent with the spectral features we observe in our Class~0 spectra, high enough to produce the conditions for CO overtone emission while not high enough to crush the stellar magnetosphere suggested by our Br~$\gamma$ emission if the protostellar magnetic field is approximately kilogauss in strength. 
%Our magnetospheric mass accretion rate estimate (6x10$^{-6}$ M\textsubscript{\(\odot\)}/yr) is consistent with the numerous spectral features we observe in our Class 0s, capable of producing the conditions for CO overtone emission while plausibly not overpowering the magnetosphere suggested by our Br $\gamma$ emission.

\item  Our H$_{2}$ line analysis (Section~\ref{subsec_ratios}) derives line ratios that argue in favor of shocks as their source excitation mechanism, similar to that of observed Class~Is. Along with the large H$_{2}$ spatial extensions and prevalence of H$_{2}$ emission we observe in our Class~0s, we interpret this as evidence of strong jets or a disk wind in the near stellar environment, consistent with earlier evidence for strong outflows in these systems on larger scales. 

%the presence of strong jets or a disk wind in the Class 0 phase. 
%3. We find the frequency of CO band emission in our observed Class 0 is notably higher than that of observed Class Is. The prevalence of emission features among our Class 0s may be evidence of an increased accretion rate for earlier (Class 0) type protostars relative to later types (Class I), with Class 0s more frequently in an "on" state of accretion. Future work including larger samples of Class 0s will help determine the prevalence of a potential underlying selection bias.

%4. Collectively, our multiple lines of evidence support the "X-wind" paradigm for YSO evolution, which predicts a strong magnetic field coupling with the circumstellar disk and produces high excitation H$_{2}$. This picture, successful at explaining many of the observed properties of Class I and IIs (\citealt{1994ApJ...426..669H}), may also apply to Class 0s. In this way, Class 0s appear to be evolving similarly to active Class I and IIs.
\end{enumerate}

Taken collectively, these results are beginning to characterize the accretion properties of Class~0s. In particular, we find Class~0 protostars may share more similarities with their Class~I counterparts than previously thought, with evidence of inner disk accretion properties similar to actively accreting Class~Is and magnetospheres being established at the earliest observable evolutionary stages. %even earlier evolutionary stages than previously known. 
These organized magnetic fields appear to be plausibly capable of mediating the accretion process, as Class~0s build towards their final mass.  

%% file: Acknowledgements.tex
\acknowledgements

This research has made use of the SIMBAD database and the VizieR catalogue access tool, operated at CDS, Strasbourg, France. This research has also made use of NASA's Astrophysics Data System and ds9, a tool for data visualization supported by the Chandra X-ray Science Center (CXC) and the High Energy Astrophysics Science Archive Center (HEASARC) with support from the JWST Mission office at the Space Telescope Science Institute for 3D visualization. Multiple Python libraries aided the analysis of our data including matplotlib, a Python library for publication quality graphics \citep{Hunter:2007}, SciPy \citep{Virtanen_2020}, NumPy \citep{harris2020array}, and Astropy, a community-developed core Python package for Astronomy \citep{2018AJ....156..123A, 2013A&A...558A..33A}. IRAF is distributed by the National Optical Astronomy Observatory, which is operated by the Association of Universities for Research in Astronomy (AURA) under cooperative agreement with the National Science Foundation \citep{1993ASPC...52..173T}. These acknowledgements were compiled using the Astronomy Acknowledgement Generator.

The authors also wish to recognize and acknowledge the very significant cultural role and reverence that the summit of Mauna Kea has always had within the indigenous Hawaiian community. We are most fortunate to have the opportunity to conduct observations from this mountain. This work was supported by a NASA Keck PI Data Award, administered by the NASA Exoplanet Science Institute. Data presented herein were obtained at the W. M. Keck Observatory from telescope time allocated to the National Aeronautics and Space Administration through the agency's scientific partnership with the California Institute of Technology and the University of California. The Observatory was made possible by the generous financial support of the W.M. Keck Foundation.

%% file: SED_fitting.tex
\section{Appendix Section A: Spectral Energy Distribution Fitting}\label{sec:SED_fitting}

\subsection{Motivation}

%\textcolor{red}{\sout{Joan: maybe start with something like" To help interpret the spectroscopy results, we made a rough estimate of the extinction to the star by modeling the SED." The text is very long though and disrupts the flow of the paper. Either shorten a lot or move details to an appendix? This section should probably go after the spectroscopy results and before the discussion.}}
Prominent star-forming regions like the Perseus and Orion complexes have been observed by multiple surveys,
% (e.g., 2MASS, \textit{Spitzer} IRAC and MIPS and Bolocam; Skrutskie et al. 2006; Fazio et al. 2004; Rieke et al.2004; Werner et al. 2004; Enoch 2006) 
recording flux measurements for many young stellar objects (YSOs) over a large range of wavelengths (1--1100$\mu$m). These measurements constitute the spectral energy distribution (SED) of an object; for Class 0s, the properties of the central source, its circumstellar disk, and its massive envelope contribute to the overall shape of the SED. Fitting this shape with an SED model can constrain numerous properties including the orientation of the disk along with the foreground extinction to the object along our line of sight.  

For the central object of a Class 0, this extinction can be significantly high, with $A_{V}$ reaching up to 70--80 magnitudes. A proper comparison of our emission line fluxes with that of the more evolved Class I and II sources requires them to first be de-reddened. In extreme cases, this can account for a large correction to the overall luminosity of the spectral feature.

However, due to degeneracies between the star, disk, and envelope properties, the shape of the SED is often not unique, with a family of SED models able to fit the observed SED. The situation is further complicated by the asymmetrical nature of most Class 0s, as the observed emission leaks out from holes in the envelope and scatters off the walls of the inner cavity. Given these limitations, we take care to not over-interpret the results, solely focused on deriving a rough estimate of the foreground extinction. This estimate is only used to very approximately de-redden our observed line luminosities as a first look comparison with that observed in Class Is (Figure~\ref{fig:LogH2lumcomp} and \ref{fig:Brylumcomp}). We note our extinction estimates are likely lower limits, given they do not quantify the envelope's (likely significant) contribution to the overall extinction. In turn, our estimated Class 0 line luminosities are likely systematically under-estimated and should be considered with care in future work. %We also note the methods used to estimate $A_{V}$ for Class Is are not as feasible for Class 0s, given the heavily embedded nature of Class 0s often lead to an absence of measurements at $J$ and $H$ wavelengths.

\subsection{Procedure}

When constructing our SEDs, we find our Class 0s have varied sampling at both near-IR and mid/far-IR wavelengths. In the near-infrared, we used \textit{J}, \textit{H}, and \textit{K}$_{s}$ from the Two Micron All Sky Survey (2MASS; \citealt{2006AJ....131.1163S}) when available, otherwise supplementing with JHK measurements from UKIDSS \citep{2007MNRAS.379.1599L} when possible. In the mid-infrared, the Infrared Array Camera (IRAC; \citealt{2004ApJS..154...10F}) and the Multiband Imaging Photometer (MIPS; \citealt{2004ApJS..154...25R}) on \textit{Spitzer} provided 3.6, 4.5, 5.8, and 8.0 $\mu$m and 24 $\mu$m photometry, respectively. We consider the reported values of this \textit{Spitzer} data from numerous sources and followup observations including \citealt{2012AJ....144...31K}, \citealt{2012AJ....144..192M}, \citealt{2015AJ....150...17R}, \citealt{2015ApJS..220...11D}, and \citealt{2017ApJS..229...28G}.

When possible, we also use data from the Infrared Spectrograph (IRS; \citealt{2004ApJS..154...18H}) on Spitzer. HOPS 32, Per-emb 21, 25, and 28 were observed by both the Short-Low (SL; 5.2–14 $\mu$m) and Long-Low (LL; 14–38 $\mu$m) modules whereas Per-emb 26 was only observed in SL. In the far-infrared, some of our Perseus sources have recorded 70, 100, and 160 $\mu$m fluxes (derived from \textit{Herschel} PACS maps in \citealt{2016A&A...592A..56M}) and, in some cases, \textit{Spitzer} MIPS 70 $\mu$m fluxes. We consider both datasets in our attempt to improve the sampling of the SED peaks. HOPS 32 has recorded Herschel PACS \citep{2010A&A...518L...2P} data at 70, 100, and 160$\mu$m. 

Similarly, some of our Perseus sources also have 450, 850 $\mu$m SCUBA (\citealt{2001ApJ...546L..49S}, \citealt{2006ApJ...646.1009K}) and 1.1 mm Bolocam \citep{2006ApJ...638..293E} measurements while HOPS 32 has 350 and 870 $\mu$m \citep{2013ApJ...767...36S} by the APEX telescope using the LABOCA and SABOCA instruments (\citealt{2009A&A...497..945S}, \citeyear{2010Msngr.139...20S},  respectively).

To fit the observed SEDs, we use the YSO SED models of \citealt{2006ApJS..167..256R}, which consists of a pre-computed grid of 200,000 total models.\ Overall, our approach adopts the recommendations laid out by \citealt{2007ApJS..169..328R}. In cases where several measurements were available at similar wavelengths, we only used the highest quality one (i.e.\ MIPS data was favored over lower-resolution IRAS measurements to avoid confusion). 

At a given wavelength, many of our sources have been observed in multiple epochs.\ In some cases, recorded fluxes (particularly in the \textit{Spitzer} IRAC channels but also 2MASS K-band and MIPS 24 $\mu$m) show significant variability, up to an order of magnitude for Per-emb 26. In these instances, we average the highest and lowest recorded flux measurements, inflating the error bars to capture the observed range of values. When applicable, this procedure grants the SED fitter more freedom to fit a variety of models for these sources with well-documented activity. 

For IRAS, far-infrared, submillimeter, and Bolocam data, a lower limit of 25\% was imposed on the flux uncertainties to account for uncertainties in their absolute calibration. For all other measurements without evidence of variability, flux uncertainties were increased to 10\% if the recorded uncertainties were smaller.

% Following the methodology in \citet{2016ApJS..224....5F}, we bin the IRS spectrum to ensure the IRS spectra 

In some cases, we find discrepancies between the flux measurements recorded at similar wavelengths but by different instruments (the overlap between IRAC channels 3 and 4 and MIPS 24 $\mu$m with our IRS spectra). This could be explained by calibration or extraction problems in the IRS spectrum (extended emission around the target or a close companion), or variability. Following the methodology in \citealt{2016ApJS..224....5F}, we assumed the former scenario if the flux deviations between IRS and IRAC and between IRS and MIPS were similar and more than 10\%, scaling the IRS spectrum to the MIPS 24 $u$m flux in those cases. We scale the IRS spectrum of Per-emb 25, Per-emb 28, and HOPS 32 by 0.816, 0.548, and 0.648, respectively. Similar to \citealt{2016ApJS..224....5F}, we also bin the IRS spectrum to 3 wavelengths after smoothing over its noisy regions to ensure it does not dominate over the photometry. Incorporating the IRS spectra at these wavelengths characterizes the 10 $\mu$m silicate feature and the mid-IR slope of the SED, which crucially constrains the resultant fits. Overall, the SEDs tend to have greater coverage at shorter wavelengths, and are thus expected to reproduce the near and mid-IR fluxes better than the sub-millimeter. 

Assumed apertures sizes for the various photometry are similar to those used in \citealt{2007ApJS..169..328R} (2\arcsec\ for UKIDSS, 3\arcsec\ for 2MASS, 5\arcsec\ for IRAC channels, 3.6\arcsec\ and 10.5\arcsec for IRS in SL mode and LL mode respectively, 10\arcsec\ for MIPS 24$\mu$m, 20\arcsec\ for MIPS 70$\mu$m, and 40\arcsec\ for SCUBA and Bolocam data). Aperture sizes of 9.6\arcsec\ for PACS 70$\mu$m, 7.2\arcsec\ for PACS 100$\mu$m, and 12.8\arcsec\ for PACS 160$\mu$m were used, given the analysis of \citealt{2016A&A...592A..56M}.

Given the estimated distance of 300 parsecs to the Persesus molecular cloud (\citealt{2018ApJ...865...73O}, \citealt{2020A&A...633A..51Z}), we adopt a distance search range of 300--340 parsecs for our Perseus sources and a search range of 420--460 parsecs (d$\sim$420 pc for the Orion Molecular Cloud, \citealt{2017ApJ...834..142K}) for HOPS 32. This range accounts for potential differences in cloud depth for our embedded sources. We adopt an $A_{V}$ extinction range of 0--100 magnitudes, assuming the extinction law of \citealt{1994ApJ...422..164K}.

\subsection{Interpretation of Best-fit Models}\label{subsec_bestfit}

% Given the SED fitting of Orion protostars in \citet{2016ApJS..224....5F}, we focus solely on our Perseus sources. 

In Figure~\ref{fig:SEDfits}, we show the final results from the SED model fitting for our Class 0s. We show the best-fit model in black and the family of models for which $\chi^{2} - \chi^{2}_{best}$ per data point $< 3$ in grey.

For the majority of our sources, we find the overall family of SED models to be well constrained and small in number. In these cases, the derived foreground extinction values have little variation, deviating by at most only a few magnitudes. In the case of Per-emb 26, however, the significant variability in the IRAC channels and MIPS 24 $\mu$m allows for a much larger number of reasonable fits along with a subsequent poor constraint on its foreground visual extinction (ranging $\sim$40 magnitudes).

Collectively, we note the best-fit models 1. capture our shortest wavelength points (those most sensitive to extinction and in the regime of the observations presented in this work) and 2. capture the depth of the crucial 10 $\mu$m silicate feature well (with the available IRS data for which only Per-emb 8 is lacking). In Table~\ref{tab:Analysistable}, we report our best-fit estimates for the visual foreground extinction to each source, noting the greater uncertainty in the case of Per-emb 26. We report additional parameters output by the SED fitting procedure in Table~\ref{tab:SEDparams}.

\section{Appendix Section B: HOPS 44 Spectrum}\label{sec:HOPS44}

Due to the low signal to noise in its spectrum, we choose to exclude HOPS 44 (the faintest object in our sample, \textit{K} $\sim$ 16.5) from the final sample of Class 0 we analyze in Section~\ref{sec:analysis}. Interestingly, we may be seeing evidence of photospheric absoprtion features in its spectrum. We show its reduced (and smoothed) spectrum in Figure~\ref{fig:HOPS44smoothspec} and encourage follow-up observations of this promising source.